\documentclass[reprint, amssymb, amsmath,aps,]{revtex4-1}
\usepackage{graphics}
\usepackage{amsmath}
\usepackage{amsfonts}
\usepackage{latexsym}
\usepackage{epstopdf}
\usepackage{bm}%
\usepackage{ulem} 
\usepackage{color}

\newcommand{\ket}[1]{\vert #1 \rangle}

\DeclareFontFamily{U}{cmbsy}{}
\DeclareFontShape{U}{cmbsy}{m}{n}{ <5> <6> <7> <8> <9> gen * cmbsy
       <10> <10.95> <12> <14.4> <17.28> <20.74> <24.88> cmbsy10}{}
\DeclareMathAlphabet{\calb}{U}{cmbsy}{m}{n}

\newcommand{\dyadic}[1]{{#1}
\setbox0=\hbox{$\scriptstyle\leftrightarrow$}
   \setbox2=\hbox{$#1$}
   \dimen0=.5\wd0 \advance\dimen0 by-.5\wd2
   \advance\dimen0 by-\wd0
   \kern\dimen0
{^{\hbox{$\scriptstyle\leftrightarrow$}}}}

\begin{document}

\title{A Multiple-Band Rydberg-Atom Based Receiver/Antenna: AM/FM Stereo Reception}
\author{Christopher~L.~Holloway}
\email{christopher.holloway@nist.gov}
\author{Matthew T. Simons}
\author{Abdulaziz H. Haddab}
\author{Joshua A. Gordon}
\affiliation{National Institute of Standards and Technology (NIST),
Boulder,~CO~80305}
\author{Stephen D. Voran}
\affiliation{Institute for Telecommunication Sciences (ITS),
Boulder,~CO~80305}


\markboth{Quantum Physics -Meets- Stereo Radio}{Rydberg-Atom Receiver}

\date{\today}

\begin{abstract}
With the re-definition of the International System of Units (SI) that occurred in October of 2018, there has recently been a great deal of attention on the development of atom-based sensors for metrology applications.  In particular, great progress has been made in using Rydberg-atom based techniques for electric (E) field metrology. These Rydberg-atom based E-field sensors have made it possible to develop atom-based receivers and antennas, which potentially have many benefits over conventional technologies in detecting and receiving modulated signals.  In this paper, we demonstrate the ``first'' multi-channel atom-based reception of both amplitude (AM) and frequency (FM) modulation signals. We demonstrate this by using two different atomic species in order to detect and receive AM and FM modulated signals in stereo. Also, in this paper we investigate the effect of Gaussian noise on the ability to receive AM/FM signals. These results illustrate the multi-band (or multi-channel) receiving capability of a atom-based receiver/antenna to produce high fidelity stereo reception from both AM and FM signals. This paper shows an interesting way of applying the relatively newer (and something esoteric) field of quantum-optics and atomic-physics to the century old topic of radio reception.

\end{abstract}

\maketitle

\section{Introduction}

Rydberg atoms are atoms with one or more electrons excited to a very high principal quantum number $n$ \cite{gal}. These atoms have several useful properties that scale as $n$. They have very large dipole moments (that scale as $n^2$). Their polarizability scales as $n^7$, and their lifetime scales as $n^3$. The spacing between the Rydberg levels scales as $1/n^3$. Rydberg atoms have large range interactions between each other that scales as $n^4/R^3$ (where $R$ is the  inter-atomic distance) and have a van der Waals interaction that scales as $n^{11}/R^6$. These various properties allow for a large array of applications and interesting physics. For example, (1) the large dipole moments make them sensitive to electrical fields, making for good field sensing, (2) the long lifetimes could lead to the development of new laser sources, and (3) the large interaction lengths create the possibilities for qubits and highly-entangled cluster states, just to mention a few.

We (and others) have made significant progress in the development of new radio frequency (RF) electric (E) field strength and power metrology techniques based on the large dipole moments associated with Rydberg states of alkali atomic vapor [either cesium ($^{133}$Cs) or rubidium ($^{85}$Rb)] placed in glass cells \cite{r2}-\cite{r20}. In this approach, we use the phenomena of electromagnetically induced transparency (EIT) for the E-field sensing. These measurements are performed either when the RF field is on-resonance of a Rydberg transition (using Autler-Townes (AT) splitting) or off-resonance (using AC Stark shifts).  These Rydberg-atom techniques allow for the development of an E-field probe that does not require a calibration (since an absolute value of the field is determined by the atomic properties of the Rydberg atom itself) and provide a self-calibrating, direct SI-traceable method for RF E-field metrology. One sensor can operate from tens of MHz to 1~THz.

This Rydberg-atom based sensor can act as a compact reciever/antenna. In particular, Rydberg atoms can lead to the development of a quantum-based receiver that measures the amplitude, phase, and polarization of modulated electric fields over a frequency range of hundreds of MHz to 1~THz. Research into using Rydberg-atoms to receive amplitude modulation (AM) and frequency modulation (FM)  signals is new.  The first demonstration of this was presented in \cite{dan}, and a few others have started investigating this topic \cite{rc1}-\cite{rc4}. This has led to the new term ``atom-radio'' coined in \cite{rc3} and \cite{atomradio}. Along with others (see below), one of the interesting possibilities of this new techniques is the ability to have a multiple-band (or multiple-channel) receiver in a single atomic vapor cell. In this paper we demonstrate the first multiple-band receiver based on Rydberg-atoms, and as such, we demonstrate the first approach to realizing an AM/FM stereo receiver with atomic vapor.


RF systems today use complex integrated circuits and metallic structures to couple, capture, demodulate and convert signals transmitted on RF electric-fields to currents and voltages. The state-of-the-art receiver technology relies on complex circuits, mixers, amplifiers, and digitizers to receive, demodulate, and decode signals. Current systems are also heavily band-limited and size-limited in comparison to what a Rydberg-atom receiver could accomplish. RF systems are also limited to frequency bands defined by the waveguide structures within them. For example, the best metrology-grade vector network analyzers operate up to around 50~GHz after which external frequency extenders are required. These only operate at discrete frequency bands such as WR-08 (90-140~GHz) and WR-05 (140-220~GHz), resulting in a large amount of equipment necessary to operate at frequencies between 1~GHz and 1~THz. Current systems also require constant calibration.

In contrast, Rydberg atoms could be used to realize a single receiver that can span the hundreds of MHz to 1~THz range, is very compact (much smaller than the RF operating wavelength), is more sensitive than current receivers, can be self-calibrated, and can be readily included in a grander quantum communications and information architecture. Over the range of hundreds of MHz to 1~THz the large dipole moments of Rydberg atoms make possible atomic antennas that can be orders of magnitude smaller than the RF wavelength and which are not governed by the Chu limit \cite{chu} as classical antennas are. At 1~GHz, classical antennas are on the order of 300~mm in size whereas a Rydberg atom is only 1~micron in size.  Furthermore, using the EIT/AT technique in preliminary tests has shown that Rydberg atoms not only respond strongly to electric fields over the 1 THz range but also inherently demodulate time-varying signals without the need for external mixers, thereby simplifying the receiver architecture.  All these attributes suggest that it is possible to make a Rydberg-atom receiver that is sub wavelength, compact, very broad band, sensitive, and which can alone achieve what currently takes many pieces of electronic equipment.

These atom-based receivers can offer advantages over current technologies: (1) nano-size antennas and receivers, (2) no Chu limit requirements as is the case for standard antennas (the atoms have the same bandwidth response over the entire frequency range of 100~MHz to 1~THz), (3) direct real-time read out, (4) no need for traditional de-modulation electronics because the atoms automatically perform the demodulation, (5) multi-band (or mutli-channel) operation in one compact vapor cell, (6) the possibility of electromagnetic interference-free receiving, and (7) ultra-high sensitivity reception from 100~MHz to 1~THz. When all said and done, the possibility of a chip-scale multi-band atom receiver.

One of the benefits of the Rydberg atoms E-field measurement technique is that it is a broadband sensor, with one sensor it is possible to measure RF E-fields from a few hundred megahertz to 1~THz. There are various ways to take advantage of this to achieve a multi-channel receiver with Rydberg atoms. One can use one atomic species (say $^{133}$Cs or $^{85}$Rb) and use different laser wavelengths (explained below) to receive different signals (or different communication channels), where each channel corresponds to a different laser wavelength. This approach will be the topic of a future publication.  The approach we will discuss here is based on using two atomic species ($^{133}$Cs and $^{85}$Rb) simultaneously, where each atomic species will detect and receive a separate communication channel. This will allow for two independent sets of data to be received simultaneously. We demonstrate this by transmitting, detecting, receiving, and playing (recording) a musical composition  in stereo. We also investigate the effect of background noise on the ability to receive the audio signal.

\section{Description of the Rydberg-atom Detection Technique}

For an AM/FM receiver/antenna, we can leverage recent work in the development of a new atom-based, SI-traceable, approach for determining E-field strengths (which is based on a novel Rydberg-atom spectroscopic approach for RF E-field strength measurements \cite{r2}-\cite{r20}).  There are various ways of explaining the concept of this E-field measurement approach (see \cite{r3}, \cite{r4}, \cite{r15} from an atomic physics viewpoint and \cite{r2} from a measurement viewpoint). Here we only give a brief explanation; see those references for more details. The concept uses a vapor of alkali atoms placed in a glass cell (referred to as a ``vapor'' cell) as a means of detecting and receiving the RF E-field or signal. The EIT technique involves using two lasers, one laser (called a ``probe'' laser) is used to monitor the optical response of the medium in the vapor cell and a second laser (called a ``coupling'' laser) is used to establish a coherence in the atomic system. When the RF E-field is applied, it alters the susceptibility of the atomic vapor seen by the probe laser as it propagates through the vapor cell. By detecting the probe light propagating through the cell (i.e., the power of the probe laser transmitted though the cell), the RF E-field strength can be determined.

\begin{figure}[!t]
\centering
\scalebox{.35}{\includegraphics*{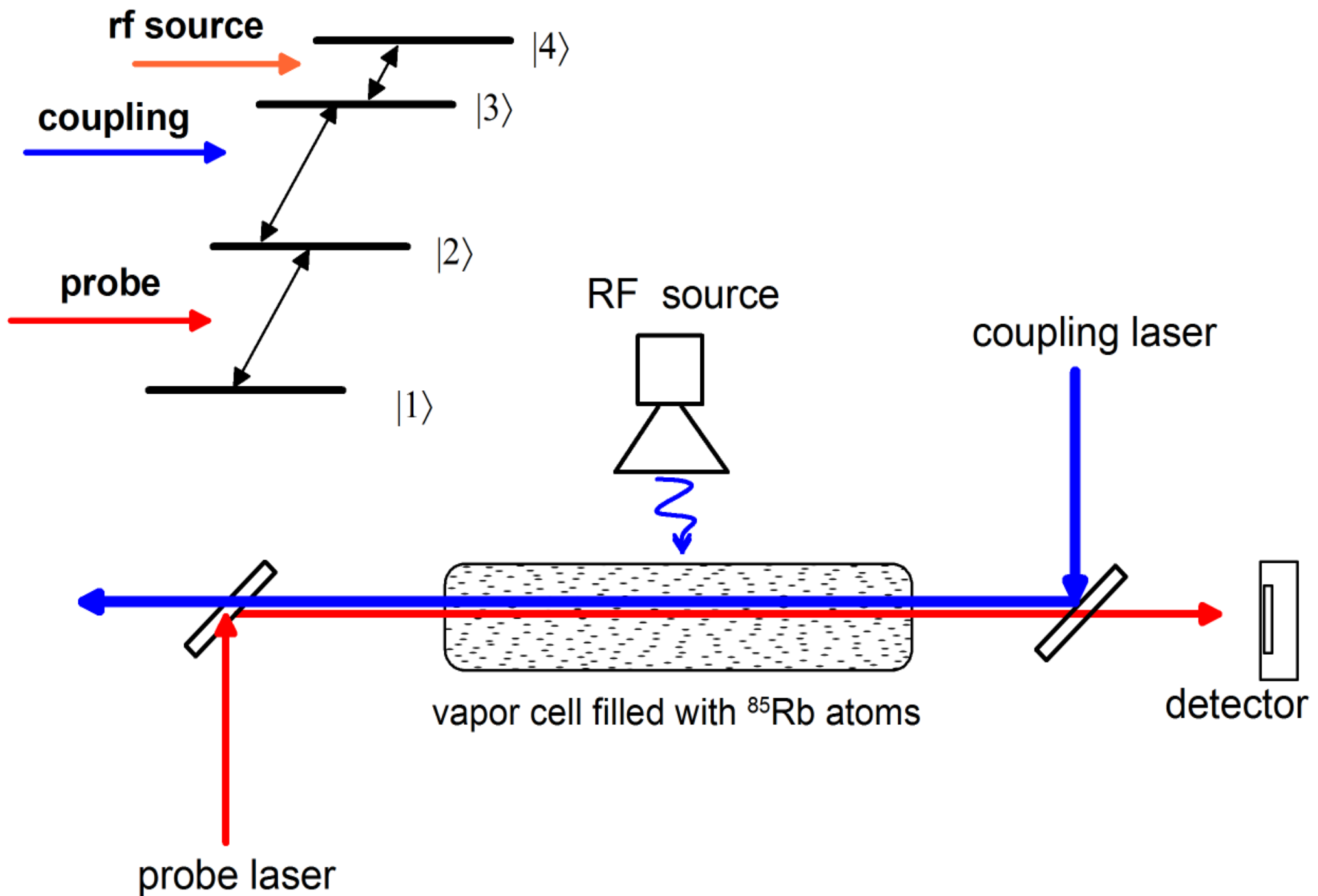}}
\caption{Illustration of a four-level system, and the vapor cell setup for measuring EIT, with counter-propagating probe and coupling beams.}
\label{4level}
\end{figure}

Consider a sample of stationary four-level atoms illuminated by a single weak (``probe") light field, as depicted in Fig.~\ref{4level}.  This figure only shows $^{85}$Rb, but $^{133}$Cs is explained in the same way. In this approach, one laser is used to probe the response of the atoms and a second laser is used to couple to a Rydberg state (the “coupling” laser). In the presence of the coupling laser, the atoms become transparent to the probe laser transmission (this is the concept of EIT). The coupling laser
wavelength is chosen such that the atom is in a sufficiently
high state (a Rydberg state) such that an RF
field coherently couples two Rydberg states (levels 3 and 4 in
Fig.~\ref{4level}).
The RF transition in this four-level atomic system causes the probe laser transmission spectrum (the EIT signal) to split.  This splitting of the probe laser spectrum is easily measured and is directly proportional to the applied RF E-field amplitude (through Planck's constant and the dipole moment of the atom). By measuring this splitting ($\Delta f_m$), we get a direct measurement of the magnitude of the RF E-field strength for a time-harmonic field from \cite{r3}, \cite{r4}, \cite{r14}, \cite{r15}:
\begin{equation}
|E| = 2 \pi \frac{\hbar}{\wp} D\, \Delta f_m= 2 \pi \frac{\hbar}{\wp}\Delta f_0 \,\,\, ,
\label{mage}
\end{equation}
where $\hbar$ is Planck's constant, $\wp$ is the atomic dipole moment of the RF transition (see Ref. \cite{r3}, \cite{r17} for discussion on determining $\wp$ and values for various atomic states), and $\Delta f_m$ is the measured splitting, $\Delta f_o=D\, \Delta f_m$, and $D$ is a parameter whose value depends on which of the two lasers is scanned during the measurement. If the probe laser is scanned, $D=\frac{\lambda_p}{\lambda_c}$, where $\lambda_p$ and $\lambda_c$ are the wavelengths of the probe and coupling laser, respectively. This ratio is needed to account for the Doppler mismatch of the probe and coupling lasers \cite{r14}, \cite{r15}. If the coupling laser is scanned, it is not necessary to correct for the Doppler mismatch, and $D=1$.

This type of measurement of the E-field strength is considered a direct SI-traceable, self-calibrated measurement in that it is related to Planck's constant (which will become an SI-defined quantity in May 2019 through the re-definition of the SI), the atomic dipole moment $\wp$ (a parameter which can be calculated very accurately \cite{r3}, \cite{r17}), and only requires a relative optical frequency measurement ($\Delta f_m$, which can be measured very accurately and is calibrated to the hyperfine atomic structure \cite{r5}).

\begin{figure}
\centering
\scalebox{.3}{\includegraphics*{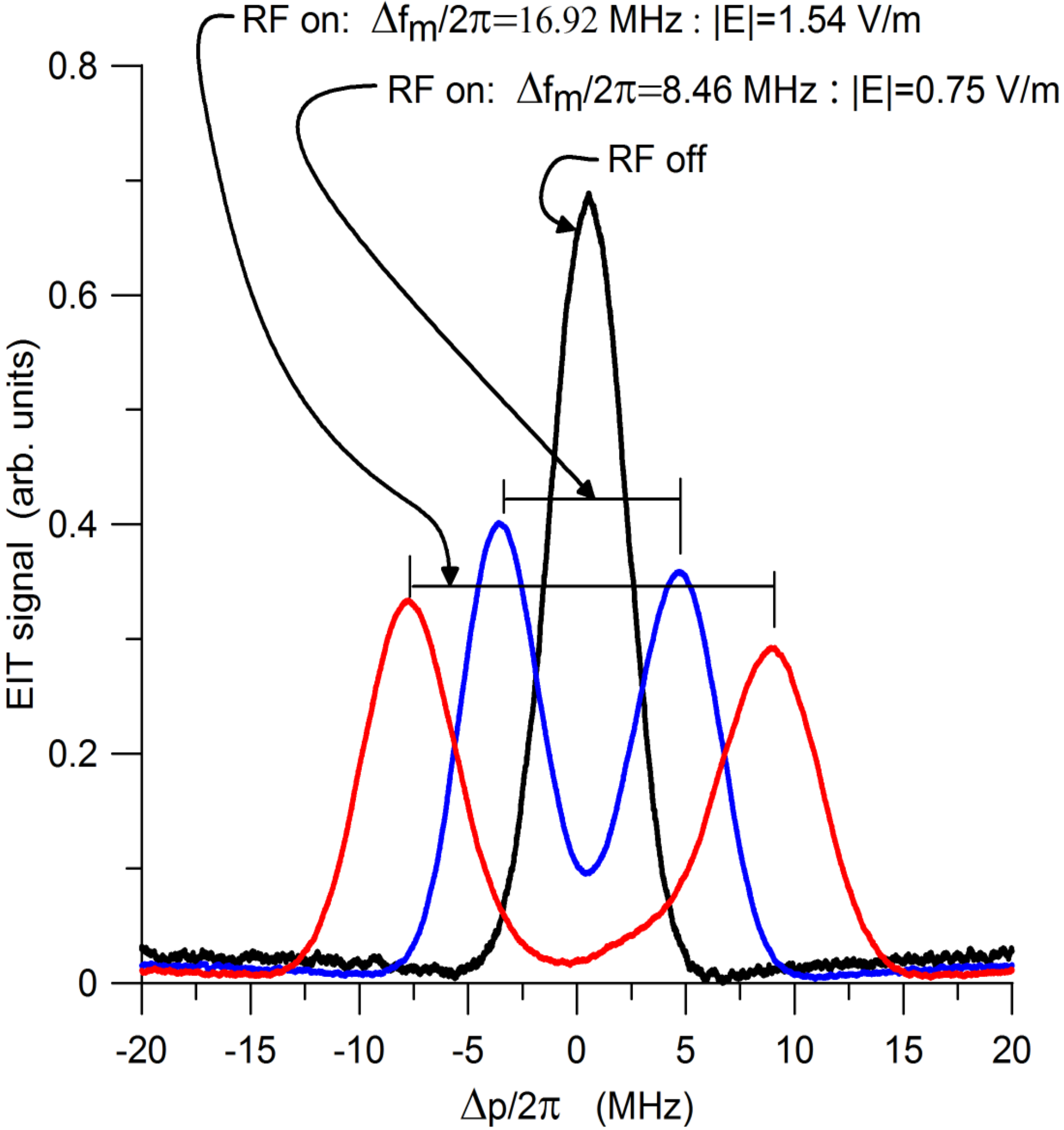}}
\caption{Illustration of the EIT signal (i.e., probe laser transmission through the cell) as a function of probe laser detuning $\Delta_p$. This dataset is for 20.64~GHz and corresponds to this following 4-level $^{85}$Rb atomic system: $5S_{1/2}-5P_{3/2}-47D_{5/2}-48P_{3/2}$.}
\label{EIT}
\end{figure}

A typical measured spectrum for an RF source with different E-field strengths is shown in Fig.~\ref{EIT}. This figure shows the measured EIT signal for two applied E-field strengths. In this figure, $\Delta_p$ is the detuning of the probe laser  (where $\Delta_p=\omega_p-\omega_o$; $\omega_o$ is the on-resonance angular frequency of the Rydberg state transition and $\omega_p$ is the angular frequency of the probe laser). Notice that the AT splitting increases with increasing applied E-field strength. To obtain these results, we use $^{85}$Rb atoms and the levels $\ket{1}$, $\ket{2}$, $\ket{3}$, and $\ket{4}$ correspond respectively to the $^{85}$Rb  $5S_{1/2}$ ground state,  $5P_{3/2}$ excited state, and two Rydberg states. The coupling laser is locked to the $5P_{3/2}$ -- $47D_{5/2}$ $^{85}$Rb Rydberg transition ($\lambda_c=480.271$~nm).
The probe laser is scanned across to the D2 transition ($5S_{1/2}$-$5P_{3/2}$ or wavelength of $780.241$~nm \cite{stackrb}). We modulate the coupling laser amplitude with a 30~kHz square wave and detect any resulting modulation of the probe transmission with a lock-in amplifier. This removes the Doppler background and isolates the EIT signal, as shown by the curve with one peak in Fig.~\ref{EIT}. Application of RF (details on one method is discussed below) at 20.64~GHz to couple states $47D_{5/2}$ and $48P_{3/2}$ splits the EIT peak as shown in the two other curves in the figure.

By measuring the frequency splitting ($\Delta f_m$) of the EIT peaks in the probe spectrum, the the E-field amplitude can be calculated from (\ref{mage}). These calculated E-field values are also shown in Fig.~\ref{EIT}. For this measurement, the dipole moment for the resonant RF transition is $\wp=1386.7064 e a_0$ (which includes a radial part of $2830.6026 e a_0$ and an angular part of $0.48989$, which correspond to co-linear polarized optical and RF fields, where $e$ is the elementary charge; $a_0=0.529177\times 10^{-10}$~m and is the Bohr radius). In order to Calculate $\wp$, one must first numerically solve the Schr$\ddot{{\rm o}}$dinger equation for the atomic wavefunctions and then perform numerical evaluation of the radial overlap integrals involving the wavefunctions for a set of atomic states \cite{gal, r3}. For a given atomic state, these numerical calculations require one to use the quantum defects (along with the Rydberg formula \cite{gal}) for the alkali atom of interest. Using the best available quantum defects \cite{gal1}-\cite{qdcs1} for $^{85}$Rb to perform a numerical calculation of $\wp$, it is believed that $\wp$  can be determined to less than $0.1~\%$, which has been verified experimentally \cite{r17}.

\subsection{Detecting an AM/FM Signal}

\begin{figure}
\centering
\scalebox{.16}{\includegraphics*{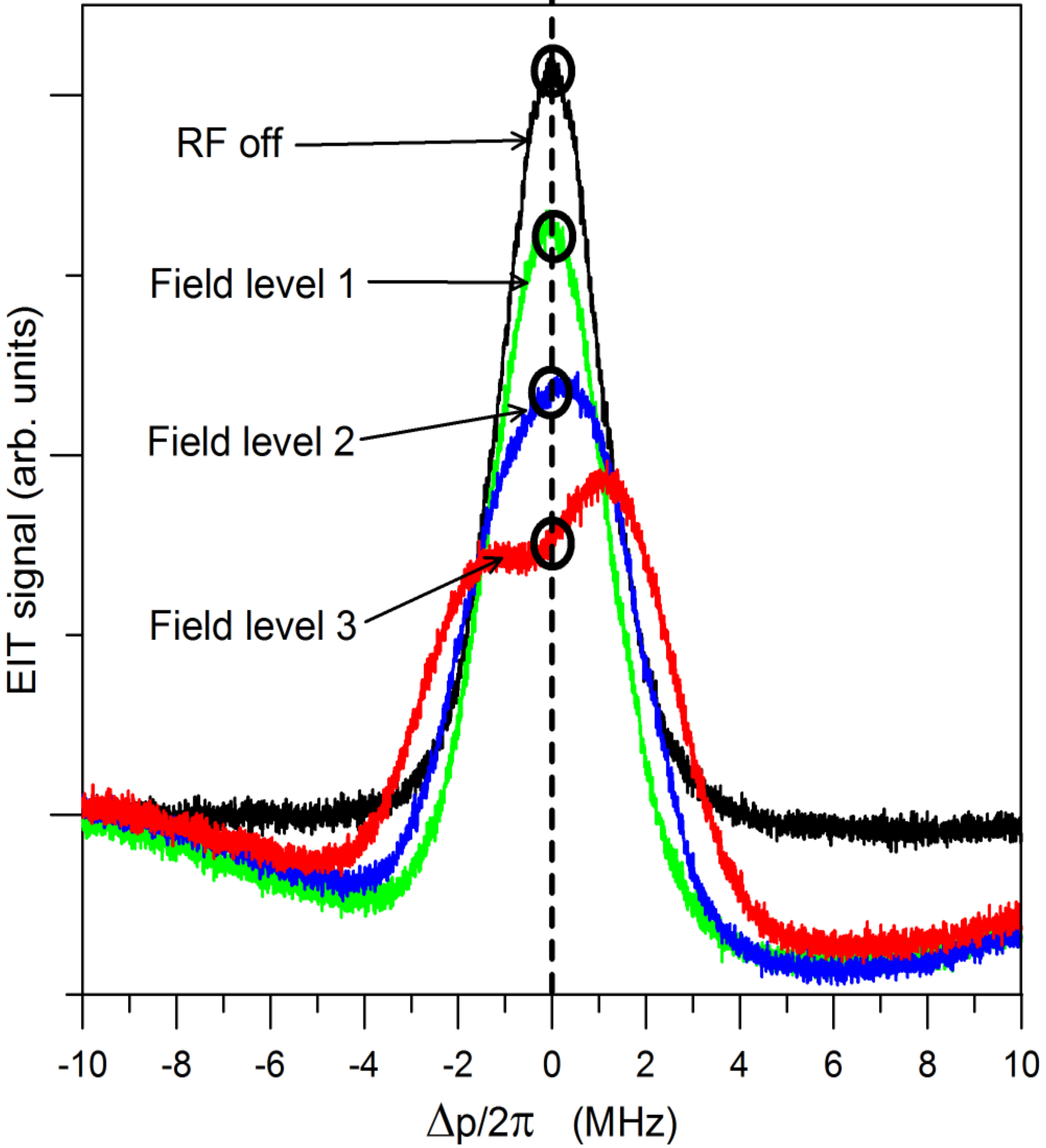}}\hspace{5mm}
\scalebox{.17}{\includegraphics*{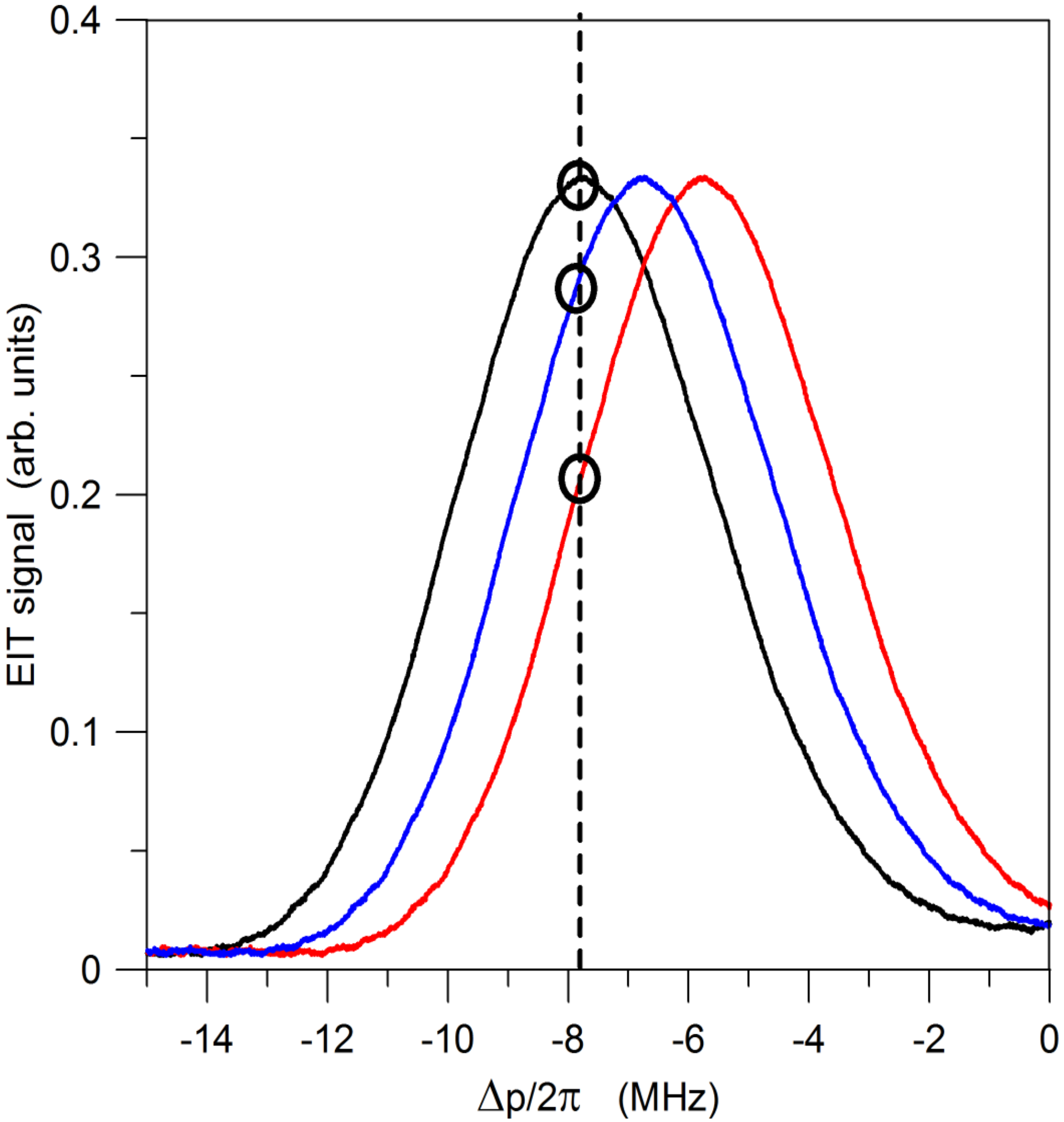}}\\
\vspace*{-2mm}
{\hspace{-1mm}\tiny{(a) \hspace{30mm} (b)}}\\
\vspace*{2mm}
\scalebox{.17}{\includegraphics*{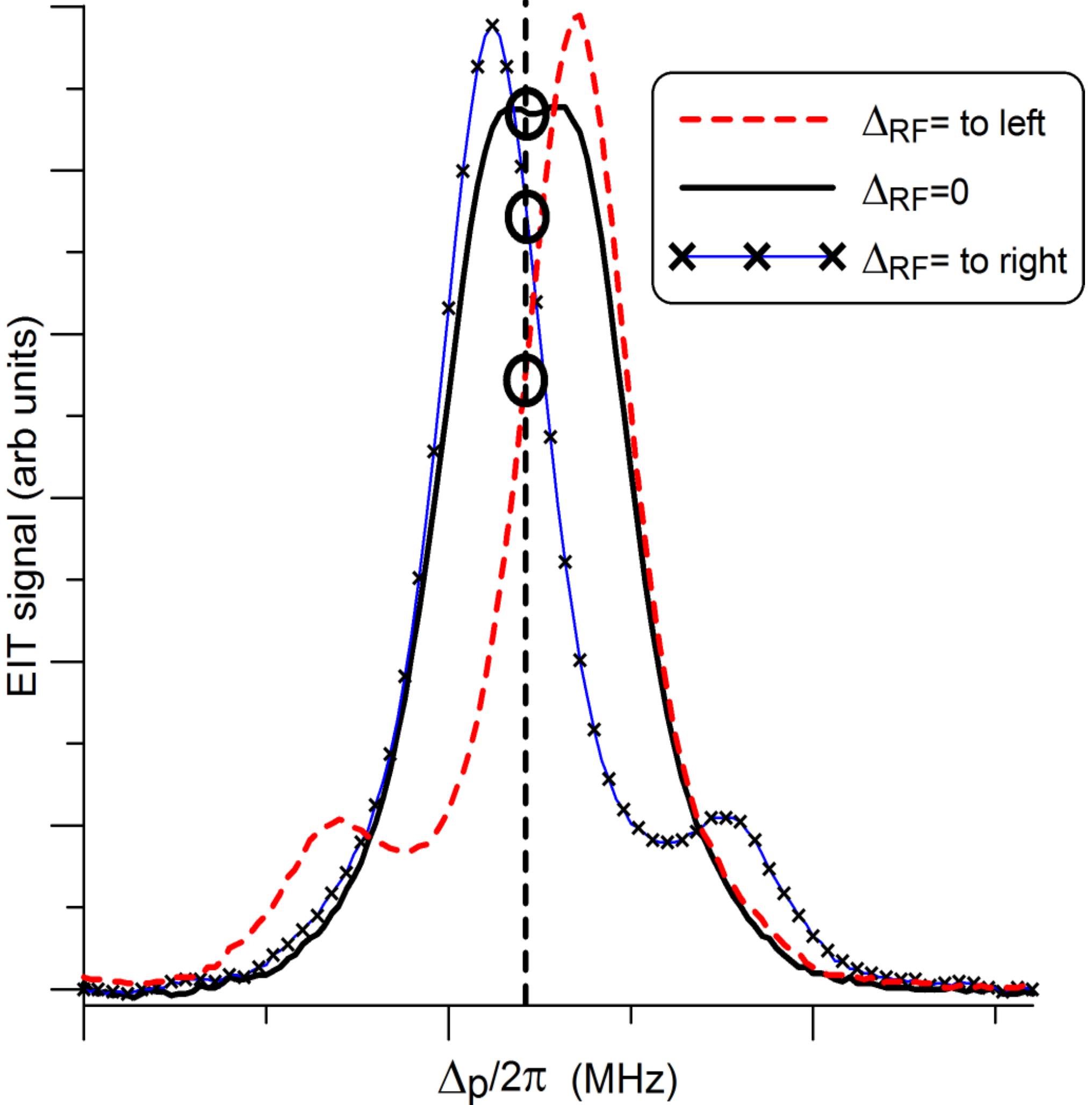}}
\scalebox{.17}{\includegraphics*{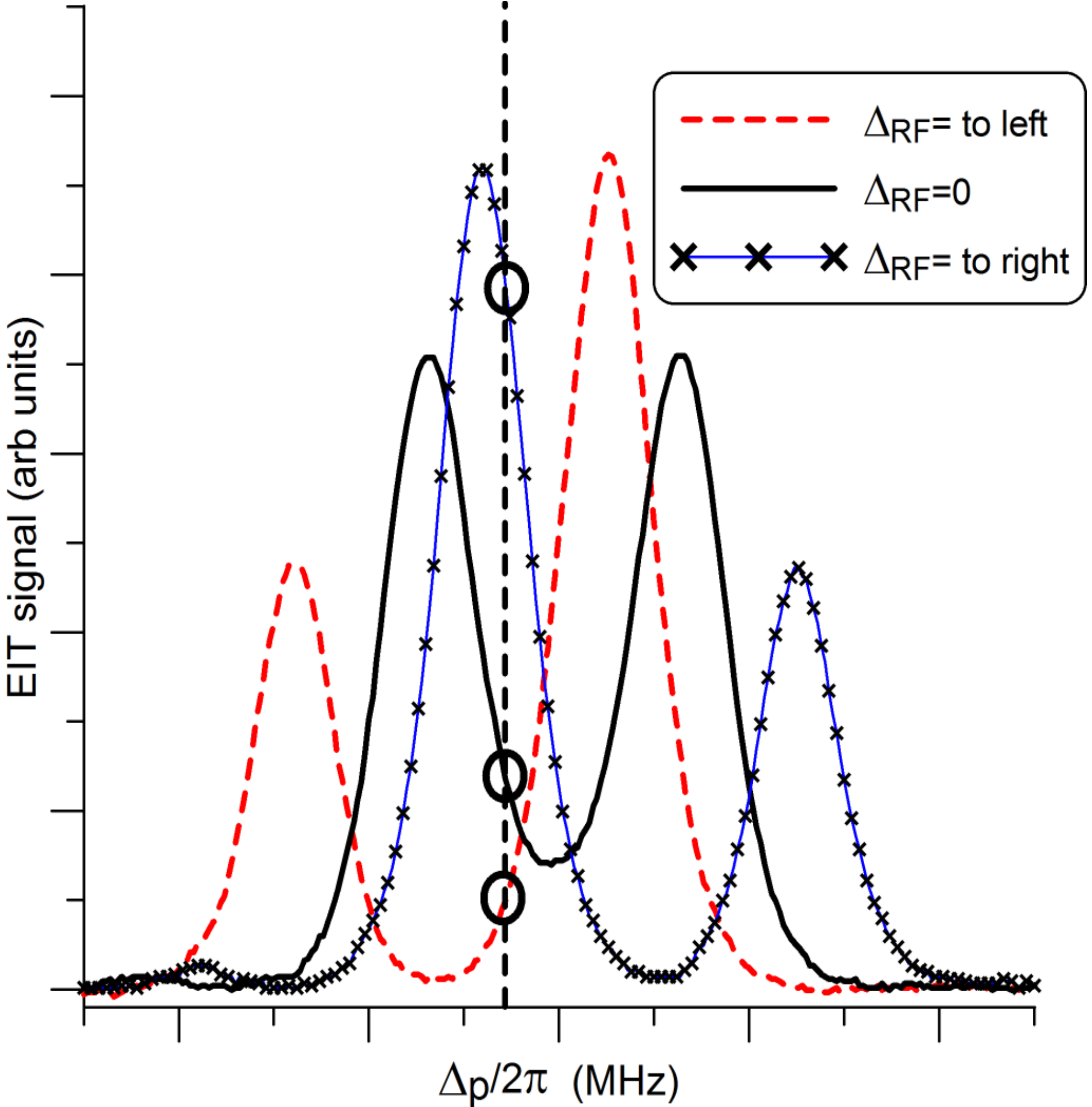}}\\
\vspace*{-2mm}
{\hspace{-1mm}\tiny{(c) \hspace{30mm} (d)}}\\
\caption{Detection scheme for an AM/FM modulated signal. (a) AM monitoring at the center of the EIT signal, (b) AM monitoring on one of the side peaks of the split EIT line, (c) FM monitoring scheme when little splitting is present, and (d) FM monitoring scheme with well defined splitting.}
\label{eitpeak}
\end{figure}

Before we discuss the detection scheme for AM/FM modulated signals, we first need to discuss the behavior of the EIT signal just before it splits. There is a minimum RF field level that is required before the splitting shown in Fig.~\ref{EIT} occurs.  When an RF field is incident onto the vapor cell and its field strength is increased from zero, the amplitude of the EIT signal decreases and its linewidth broadens before the EIT signal starts to split.  This is illustrated in Fig.~\ref{eitpeak}(a). Shown here is the EIT signal with no RF  field, and three cases for different RF field strengths. Notice that the curve labeled ``Field level 3'' corresponds to a field strength high enough to cause splitting.

Here, we discuss two possibilities for the detection of both AM and FM modulated signals. One is based on detection of the EIT signal at $\Delta_p=0$ and one is based on detection at $\Delta_p\ne0$. They both work basically the same way, and choosing one versus the other depends on the power in the RF carriers and the modulation depth used.  Recall that an AM signal is basically a carrier where the amplitude of the carrier changes. Let us start by discussing the situation where the amplitude of the carrier (and the modulation depth) is such that no splitting in the EIT signal will occur. The AM modulated carrier will only cause the peak of the EIT line to move up and down the dashed line shown in Fig.~\ref{eitpeak}(a). Therefore, if the probe laser is locked to $\Delta_p=0$ (while also locking the coupling laser to the $5P_{3/2}$-$47D_{5/2}$ Rydberg transition), the voltage output of the photo-detector (used to measure the probe laser transmission) would be directly correlated to the modulating signal.  That is, no demodulation circuity is needed, the Rydberg atoms (through the probe transition through the cell) automatically demodulate the signal, and we get a direct read-out of the baseband signal.

When the carrier frequency signal strength is large enough to split the EIT line into two peaks, AM would cause the EIT peaks to move left-to-right (or right-to-left).  This in illustrated in Fig.~\ref{eitpeak}(b).  This shows (on a zoomed-in x-axis) the peak of one of the EIT lines (the one to the left) for different values of the amplitude of the carrier signal.  Here again, if the probe laser is locked to $\Delta_p=2\pi\cdot8$~MHz (while also locking the coupling laser to the $5P_{3/2}$-$47D_{5/2}$ Rydberg transition), the voltage output of the photo-detector (see the EIT signal strength along the dashed line) would be directly correlated to the modulating signal (i.e., the atom's response basically modulate the photo-detector signals). We could just as well lock the probe laser to a wavelength just off the peak when determining the modulated signal.

The detection of a FM modulated signal works in a similar manner.
When an RF field is detuned (i.e., the RF frequency is changed) from its resonant RF transition frequency it has two main effects on the observed splitting of the EIT signal which are discussed in detail in \cite{r16}, see Figs.~\ref{eitpeak}(c) and \ref{eitpeak}(d). First, the two peaks of the EIT signal are non-symmetric (i.e., the heights of the two peaks are not the same). The second effect of RF detuning is that the separation between the two peaks increases with RF detuning.  If the probe laser is locked to some $\Delta_p$ (while also locking the coupling laser to the $5P_{3/2}$-$47D_{5/2}$ Rydberg transition), the voltage output of the photo-detector (see the EIT signal strength along the dashed line) would be directly correlated to the modulating signal (i.e., the atom's response basically modulates the photo-detector signals).

\section{Experimental Setup for AM/FM Stereo Receiver}

\begin{figure}[!t]
\centering
\scalebox{.25}{\includegraphics*{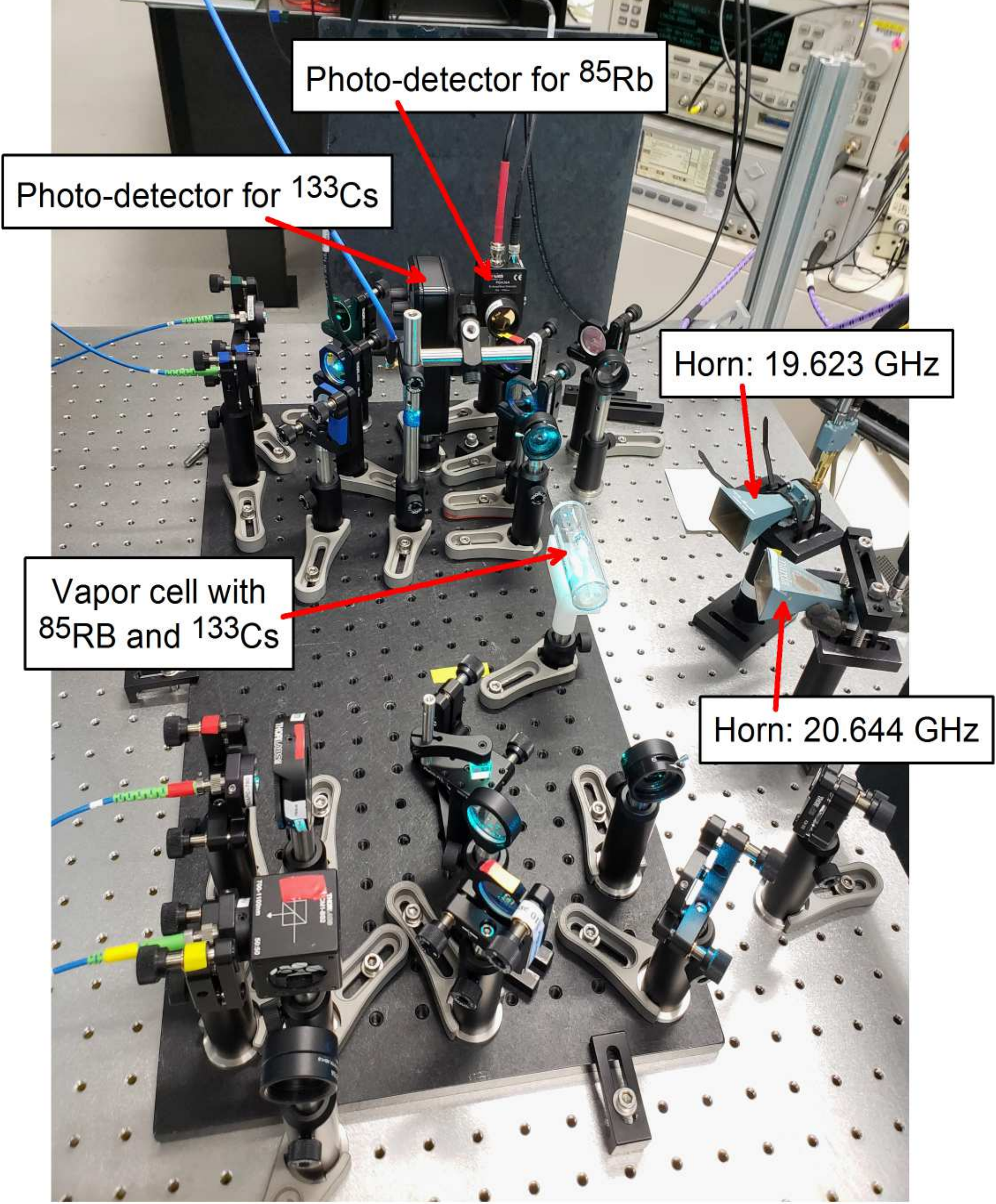}}\\
\vspace*{1mm}
{\tiny{(a)}}\\
\vspace*{2mm}
\scalebox{.34}{\includegraphics*{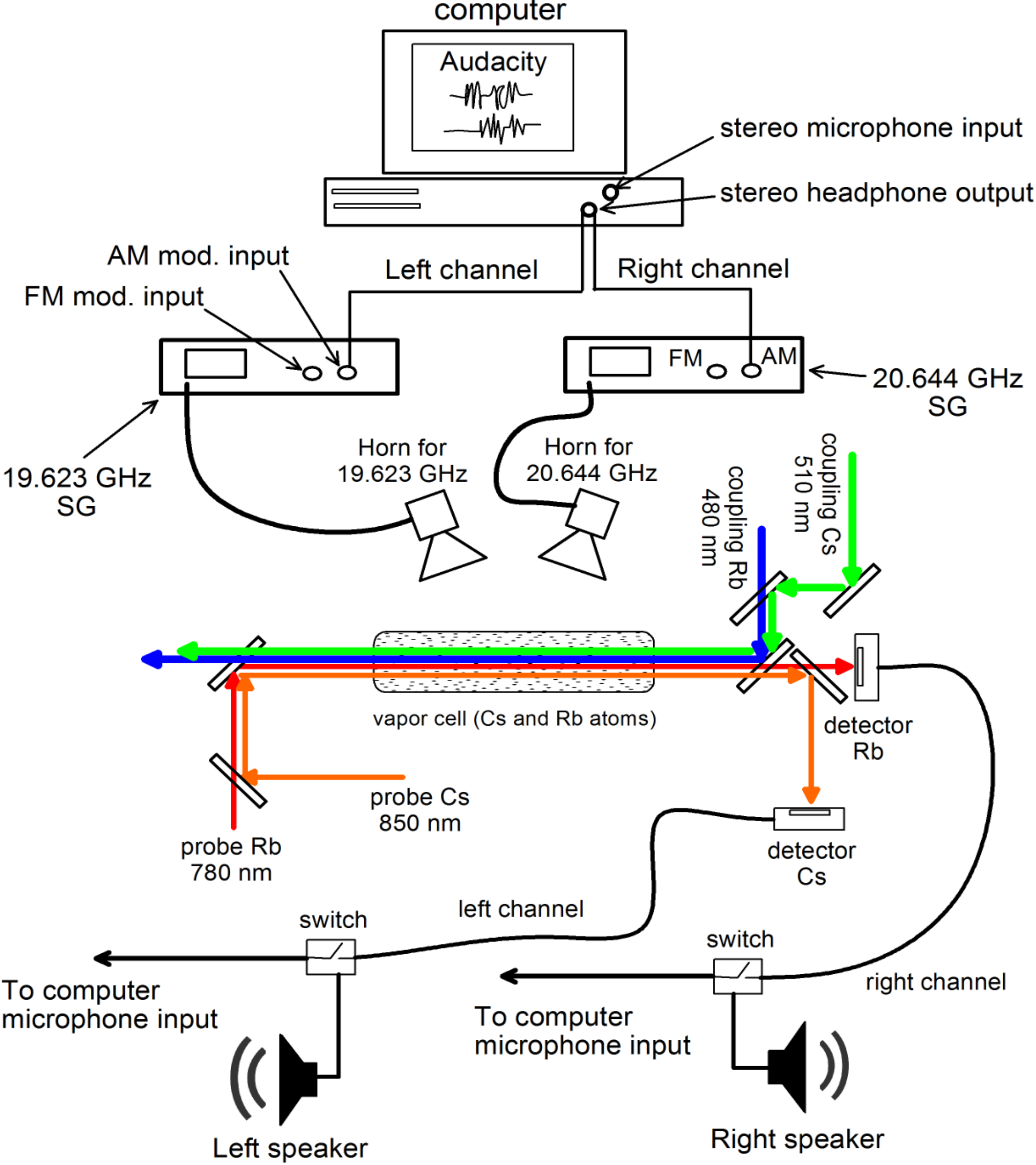}}\\
\vspace*{1mm}
{\tiny{(b)}}\\
\caption{Experimental setup for AM/FM receiver measurements using EIT: a) picture of the setup, (b) block diagram of the setup including vapor cell setup for two atomic species and four lasers (two  counter-propagating probes beams and two coupling beams.}
\label{setup}
\end{figure}

\begin{figure*}
\centering
\scalebox{.27}{\includegraphics*{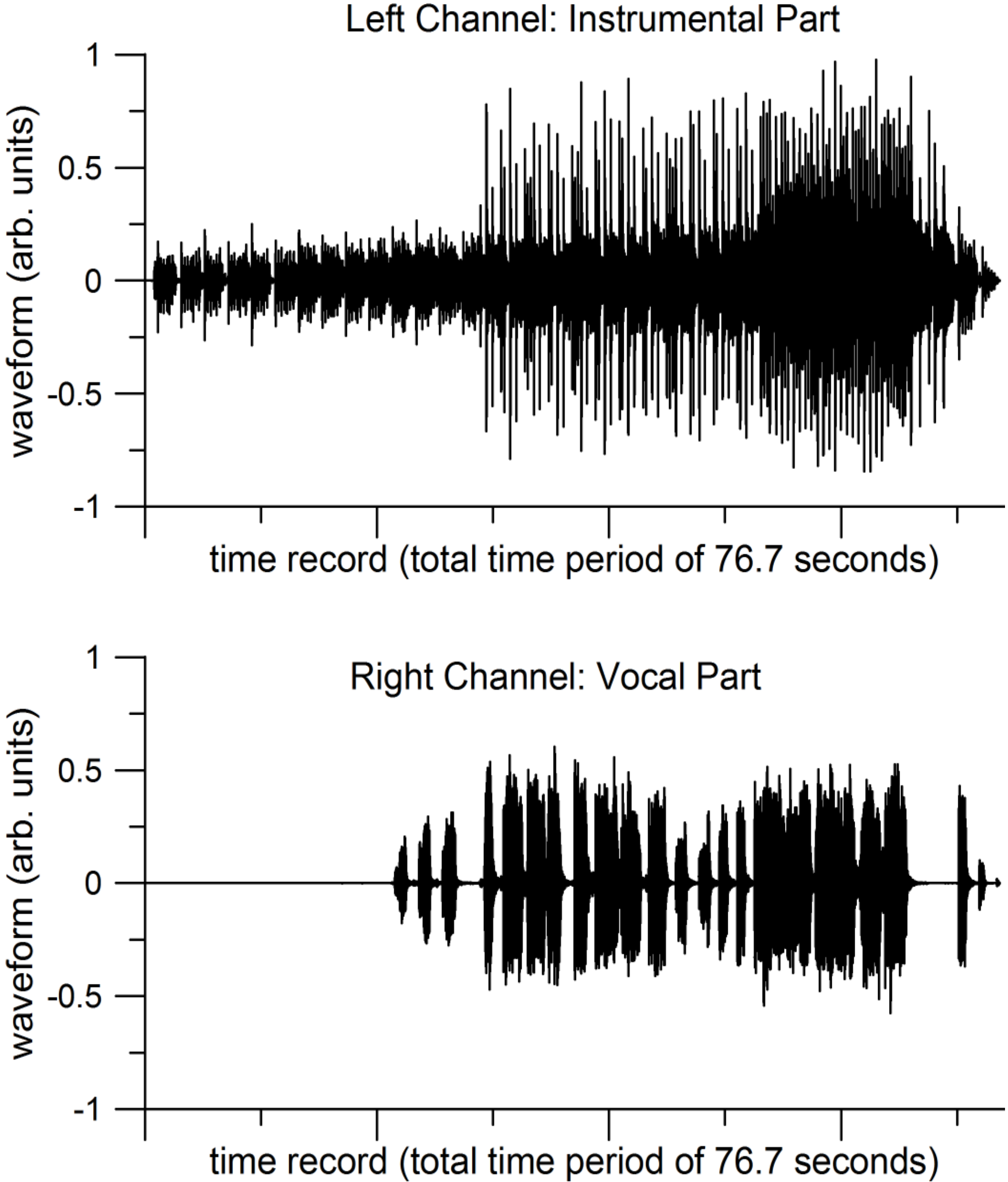}}\hspace{5mm}
\scalebox{.27}{\includegraphics*{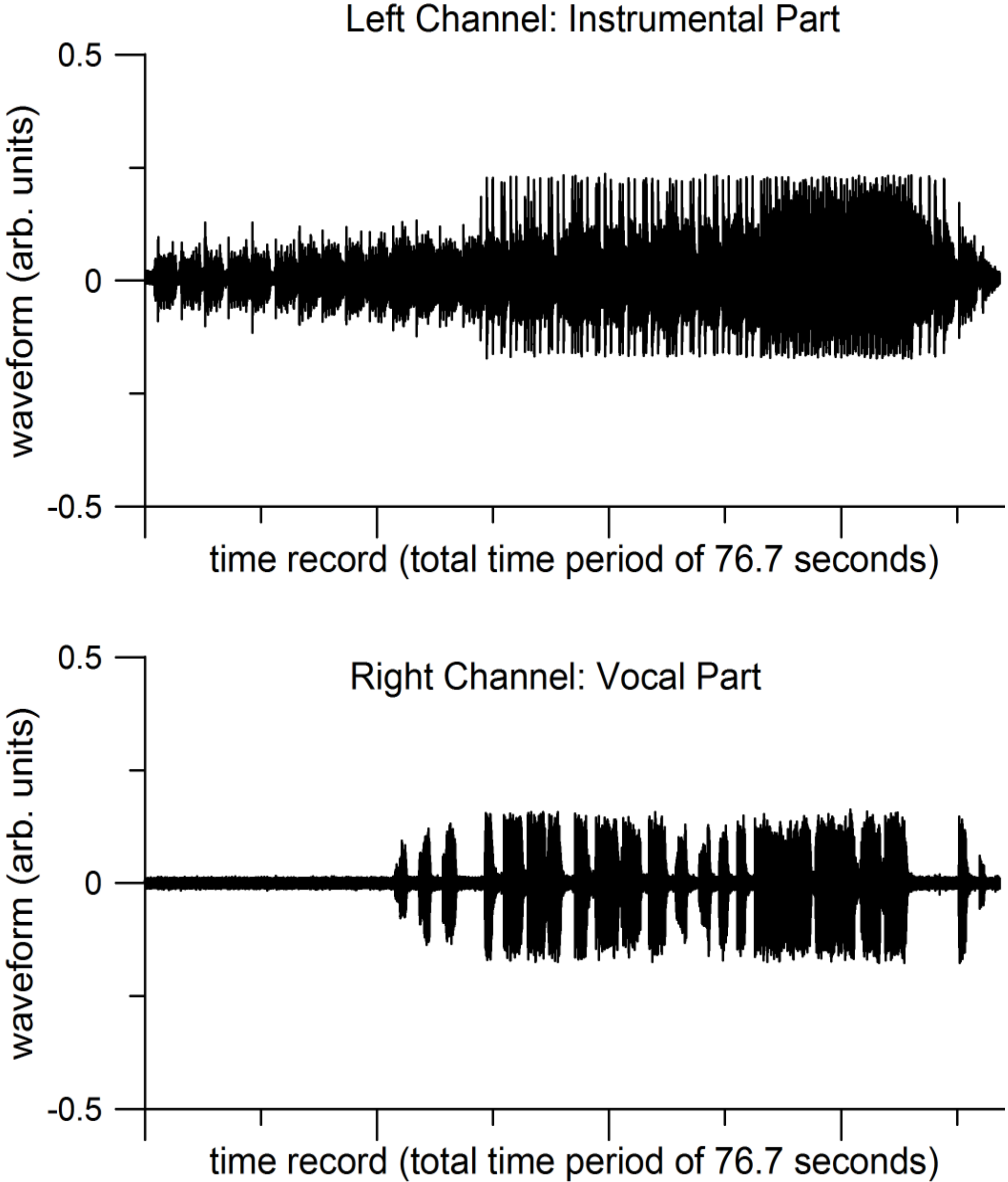}}\hspace{5mm}
\scalebox{.27}{\includegraphics*{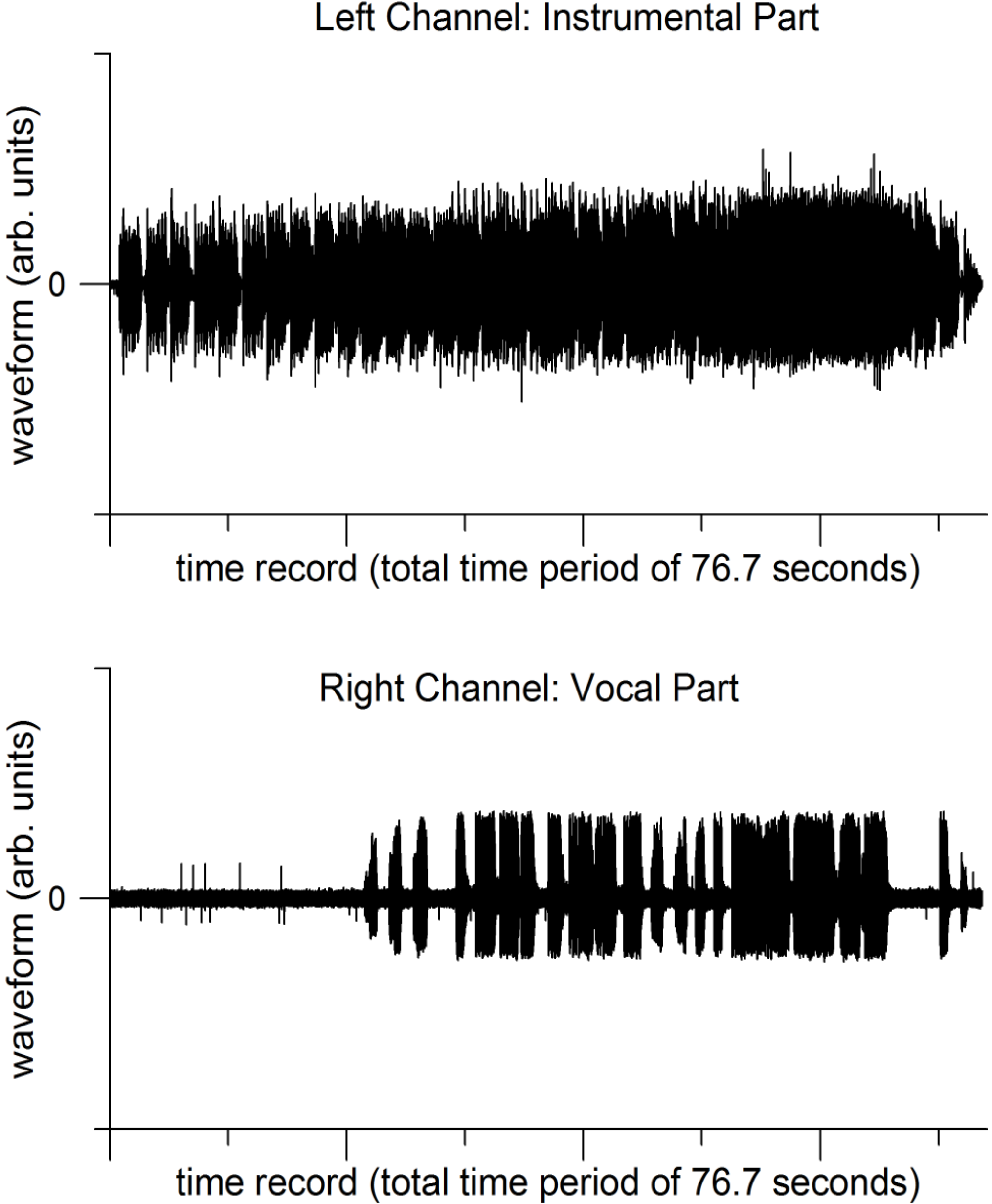}}\\
\vspace*{-2mm}
{\tiny{(a) \hspace{57mm} (b) \hspace{57mm} (c)}}\\
\caption{(a) Waveform of the musical composition used for the stereo AM/FM receiving experiments, (b) received waveform of the musical composition from the AM modulation scheme for to the atom-based stereo receiver, and (c) received waveform of the musical composition from the FM modulation scheme for to the atom-based stereo receiver.
 The top curves are the instrumental part (which we designate as the left channel), and the bottom curves are the vocal part (which we designate as the right channel). The total time period of these waveforms is approximate 76.7~seconds.}
\label{score}
\end{figure*}

The experimental setup for transmitting, detecting, receiving, playing (through a set of speakers), and recording a musical composition in stereo is shown in Fig.~\ref{setup}. We first discuss the AM scheme.
We chose a musical composition as the data to transmit and receive via the multi-channel Rydberg-atom receiver. The musical composition we chose, has both an instrumental part and a vocal part, see Fig.~\ref{score}(a). We separated these two parts into two different audio data files and saved them in the ``wav'' format. We use the open-source program {\it Audacity} (mentioning this product does not imply an endorsement, but serves to clarify the software used) to play these two audio files. The instrumental part was put on the ``left'' channel of the stereo headphone jack of the computer and the vocal part was put on the ``right`` channel of the headphone jack of the computer. The output of the headphone jack is simply a voltage waveform with a range of $\pm1$~V. These two voltage waveforms were used to modulate two different carrier frequencies. The ``left'' modulated a 19.626~GHz carrier and the ``right'' channel modulated a 20.644~GHz.  We used two different signal generators (SG) to generate these two different continuous wave (CW) signals.  The modulation was performed in two different ways, by either using the internal AM/FM modulation feature in the SG or using an external mixer. Both modulation schemes work equally well for this application, since the waveform for an audio file is limited to about 20~kHz. The SG AM and FM modulation feature is limited to 100~kHz, so for modulation rates greater than 100~kHz, the external mixer must be used (as was done in \cite{biterror} for receiving pseudo-random bit streams at different modulation rates).  The results presented below are from using the internal modulation feature of the SG.

The output from each SG was connected to two Narda 638 standard gain horn antennas (mentioning this product does not imply an endorsement, but serves to clarify the antennas used). Each antenna was placed 30~cm from a cylindrical glass vapor cell of length 75~mm and diameter 25~mm containing both $^{85}$Rb and $^{133}$Cs atomic vapor, see Fig.~\ref{setup}(a) and (b).  The two atom species require the use of four lasers. The four laser setup shown in Fig.~\ref{setup}(b) was used to detect and receive the modulated signals. The $^{85}$Rb atoms are used to receive the 20.644~GHz modulated carrier, and the $^{133}$Cs atoms are used to receive the 19.626~GHz modulated carrier.  The probe laser for $^{85}$Rb is a 780.24~nm laser focused to a full-width at half maximum (FWHM) of 750~$\mu$m, with a power of 22.3~$\mu$W. To produce an EIT signal in $^{85}$Rb (using the atomic states given in Fig.~\ref{EIT}), we apply a counter-propagating coupling laser (wavelength $\lambda_c=480.271$~nm) with a power of 43.8~mW, focused to a FWHM of 250~$\mu$m.  The probe laser for $^{133}$Cs is a 850.53~nm laser ($6S_{1/2}$-$6P_{3/2}$) focused to a full-width at half maximum (FWHM) of 750~$\mu$m, with a power of 41.2~$\mu$W. To produce an EIT signal, we couple to the $^{133}$Cs $6P_{3/2}$ -- $34D_{5/2}$ states by applying a counter-propagating coupling laser at $\lambda_c=511.1480$~nm with a power of 48.7~mW, focused to a FWHM of 620~$\mu$m. We apply an RF field at 19.626~GHz to couple states $34D_{5/2}$ and $35P_{3/2}$.

Two different photo-detectors were used to detect the transmission for each probe laser through the vapor cell (one for $^{85}$Rb and one for $^{133}$Cs). The output of the photo-detectors is a voltage waveform, and the output was connected in two different configurations. The first configuration consisted of simply connecting the output of the photo-detectors to a set of computer speakers: (a) the output for the $^{133}$Cs probe laser photo-detector was connected to left computer speaker, and (b) the output for the $^{85}$Rb probe laser photo-detector was connected to right computer speaker. In the second configuration, the output of the two photo-detectors were connected to a stereo jack and plugged into the microphone input of a computer.  We then used {\it Audacity} to record the the left and right channels separately from this microphone input.

\section{Experimental Results}

Before we performed stereo measurements, we needed to ensure the system was working correctly. This was done by modulating the two different carriers with the same waveform and ensuring that the same baseband signal could be received simultaneously by the two different atomic species (i.e., receiving on both $^{85}$Rb and $^{133}$Cs simultaneously). To accomplish this, we played an internet-based radio station on the computer and connected that signal (through the computer stereo headphone jack) to the AM input of two SGs (note the music was in mono with the same voltage waveform on the left and right channels).  We first observed the output by listening to the left and right speaker, which we observed that while there is some noise in the sound (more on this below) the left and right speaker outputs were essentially the same.  We then recorded the music (via the microphone input and with {\it Audacity}).  The output of the two channels is not shown here, but they are virtually the same. We used this approach to stream and listen to music for several hours at a time as we worked in laboratory, illustrating the long term stability of the approach.

\begin{figure}
\centering
\scalebox{.27}{\includegraphics*{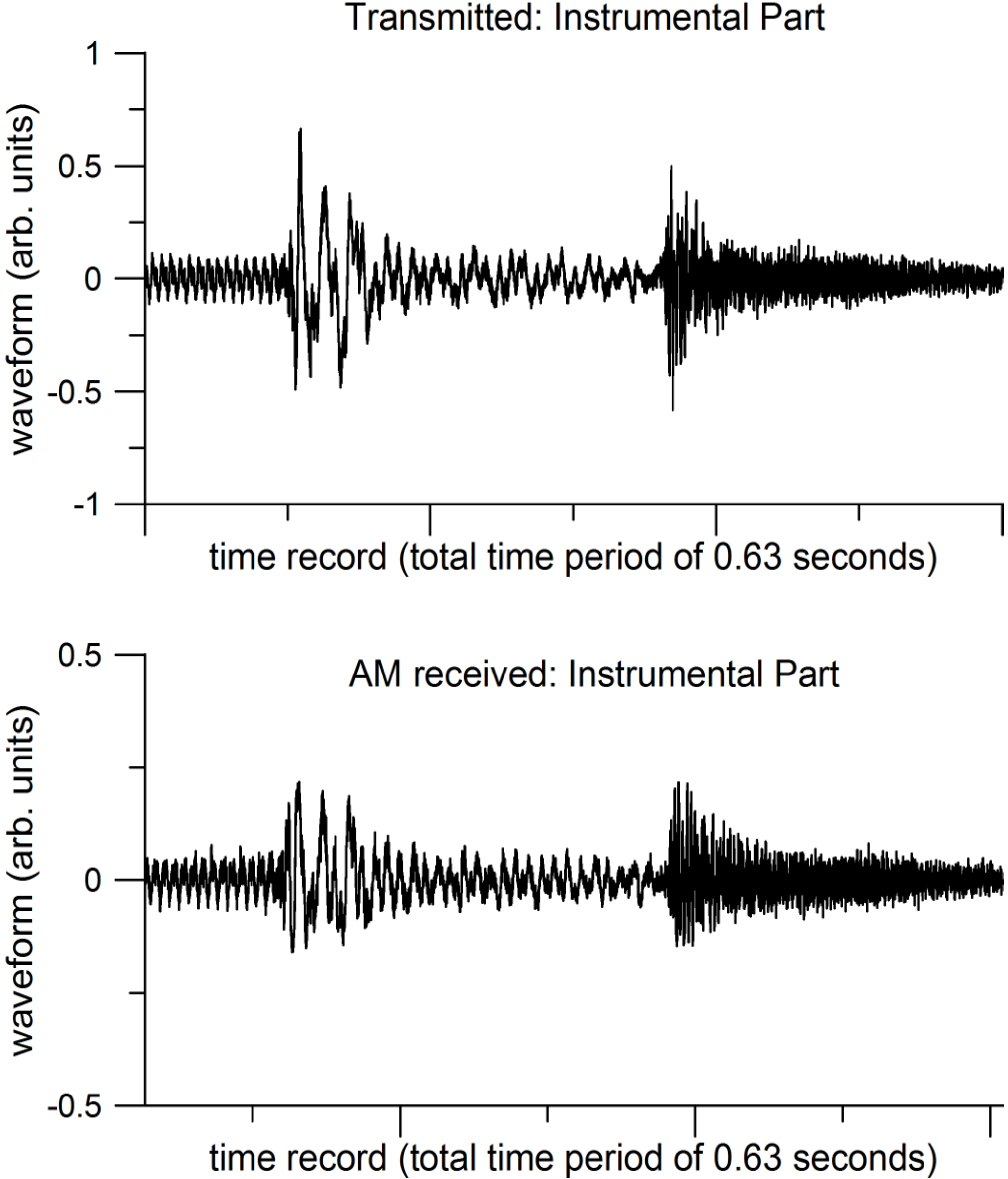}}
\caption{Comparison of a 0.63-second segment of the instrumental part of the musical composition.}
\label{compare}
\end{figure}

To illustrate stereo reception, we played the instrumental part of the musical composition shown in the top curve of Fig.~\ref{score}(a) through the left channel (which modulated the 19.626~GHz carrier) and we played the vocal part shown in bottom curve of Fig.~\ref{score}(a) through the right channel (which modulated the 20.644~GHz carrier).  In turn, the baseband of the 19.626~GHz carrier (i.e., the left channel which contains the instrumental part) was received by the $^{133}$Cs atoms, and the baseband of the 20.644~GHz carrier (i.e., the right channel which contains the vocal part) was received by the $^{85}$Rb atoms.  The output of the photo-detectors for the left and right channels were then connected to the right and left computer speakers and stereo reception was achieved (and we listened to the musical composition). The output sound from the speakers was of high fidelity, in that the musical composition was clearly audible and very understandable, although a little noise was audible (see below for discussion on audio quantity assessment). While noise was present, it had a very minor effect on the quality of the sounds. We then connected the output of the photo-detectors to the microphone jack of the computer, and recorded the two channels with {\it Audacity}.  These two recordings are shown in Fig.~\ref{score}(b).  Compared to the original data file shown in Fig.~\ref{score}(a), these are basically the same files (besides the fact that the data in Fig.~\ref{score}(b) has less amplitude than those in Fig.~\ref{score}(a)). In Fig.~\ref{score}(b) we do see some clipping (or possible amplitude compression) in the left channel (i.e., the instrumental part). While it is seen in the figure it was not apparent when listening to the musical composition. To further illustrate this point, Fig.~\ref{compare} shows a 0.63-second segment of the waveforms shown in Fig.~\ref{score}, in which we compare the transmitted waveform to that received with the AM scheme. We see that even with the clipping, the waveforms are similar (accept for the difference in amplitudes). It is believed that the clipping is due in part to the response of the photo-detector used for the left channel.  Assessing the quality of audio files is not a trivial task. For digital data one can assess the bit-error-rate (as is done in \cite{biterror} for an AM/FM atom-based receiver), but such a method can not be used on audio data (the next section does present one method to assess audio equality). With that said, these results illustrated the multi-band (or multi-channel) receiving capability of a small single vapor cell.

We next demonstrate the FM scheme by using the computer headphone outputs as the inputs to the FM feature of two SGs (using the FM modulation features in the SG).  When listening to the output of the two speakers, we notice that while there is a little more noise than observed in the AM scheme, high-fidelity music was present. Fig.~\ref{score}(c) shows received waveform from the atoms using the FM scheme.

We believed that the majority of the noise in the data sets for both AM and FM is from laser noise. While the detection could be improve and laser noise reduced, the results here illustrate the capability of an atom-based multi-channel receiver.

%

\subsection{Effects of White-Gaussian Noise}

For this to be a useful technique it is important to understand the
influence of white Gaussian noise (WGN). In fact, it is believed that this atom-based approach may be less susceptible to noise. This is confirmed in \cite{r20}, where we performed experiments measuring CW E-field strengths using this atom-based approach in the presence of band-limited white Gaussian noise (BLWGN) and we showed that the E-field strength could be detected in low {\it CW-signal to noise-power ratio} (CSNR) conditions. In this section, we report on some primary experiments to investigate how noise effects the reception of AM signals.

For this investigation we use the right channel of the musical composition (i.e., the vocal part of the composition) obtained from the AM scheme. This part is particularly useful because it has gaps of near silence and these gaps provide good locations to measure any noise that has been added.

\begin{figure}
\centering
\scalebox{.35}{\includegraphics*{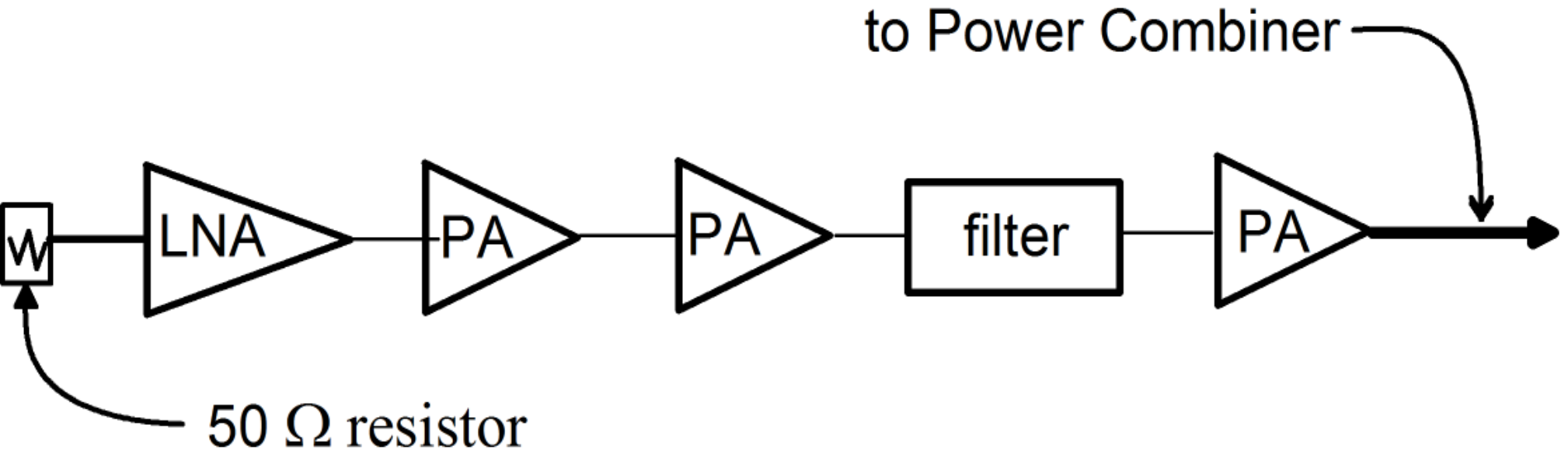}}\\
\caption{Block diagram of the noise source setup.}
\label{noise}
\end{figure}

In these tests we injected BLWGN and a CW carrier into a horn antenna.  A power combiner was connected to the input of the horn
antenna to combine the noise signal and the CW 20.644~GHz signal from a SG, such that both noise and CW signals were incident
on the vapor cell simultaneously.  The noise signal was generated by connecting a 50~$\Omega$ resistor to a series of amplifiers, as shown in Fig.~\ref{noise}. The resistor was connected a low-noise amplifier (LNA) with a gain of 27~dB then to two power amplifiers (PAs), each with a gain of 26~dB. The output of the second PA was sent to a
bandpass filter that was changed to the different bands
during the experiment. The output of the filter was then fed
into a third PA with 30~dB gain. The output of the third
amplifier was connected to the power combiner.

In these experiments we used three different filters, each with a bandwidth of ~1 GHz, with different center frequencies as
follows: Filter 1 was $\approx$20.7~GHz, Filter 2 was $\approx$19.7~GHz, and Filter 3 was $\approx$18.7~GHz. The noise power spectral densities resulting from these bandpass filters, measured with a spectrum analyzer connected to the output of the power combiner (i.e., the input to the horn antenna) are shown in \cite{r20} (see Fig.~5 therein). Using a power meter, we measured the integrated noise power (total power over the filter bandwidth) for each filter, measured at the output of the power combiner that feeds the horn antenna. The integrated power was $0.40$~dBm for Filter~1, $0.55$~dBm for Filter~2, and $-1.10$~dBm
for Filter~3.

To vary the noise levels during the experiments we added one of three different attenuators between the noise source and the power combiner. This provided a total of four different BLWGN levels for each of the three filters (one without an attenuator, and one for each of three different attenuators).
The CSNR (defined as the ratio of the CW power in the carrier to the integrated noise power, both measured at the input to the horn antenna) for these different combinations are shown in Table~\ref{t2}. During the experiment the 20.644~GHz carrier at the output of the power combiner was $-22.0$~dBm.

Across the thirteen different cases (no noise added and four added noise levels each crossed with three filters) the vocal signal is never noticeably distorted. But the noise present in the received audio signal ranges from minor to severe. We measured the noise levels in the silent intervals of the received audio signal and report the results here.

Measuring noise levels involves comparing the recorded audio files with the original transmitted audio file.
The first step is to use correlation to find and remove the time shift between the transmitted audio file and each received audio file. We then segmented each file into groups of $N=1024$ audio samples (called frames) and calculated the power in dB for each frame. We used an audio sample rate of 48,000 smp/s, so each frame has a duration of 21.3 ms.
The power of the $i^{th}$ frame is given by
\begin{equation}
 P_i=10\,\,{\rm log}_{10}\,  \sum_{j=1}^N  x^2_{(i-1)N+j}\,\,\, ,
\label{e2}
\end{equation}
where $x_k$ represents the $k^{th}$ audio sample.

We produced frame power histories for each of the thirteen cases, and for the original transmitted audio file as well.
We normalized each history to have a power of 0 dB at frame 573.
This is the frame of the transmitted audio file that has the greatest power (and each of the other thirteen signals have
maximal frame power at frame 573 as well).
This normalization step is equivalent to matching the levels of each of the received audio signals with the level of the transmitted audio signal. This is required in order to make meaningful comparisons between the noise levels associated with each of the signals.

\begin{figure}
\centering
\scalebox{.18}{\includegraphics*{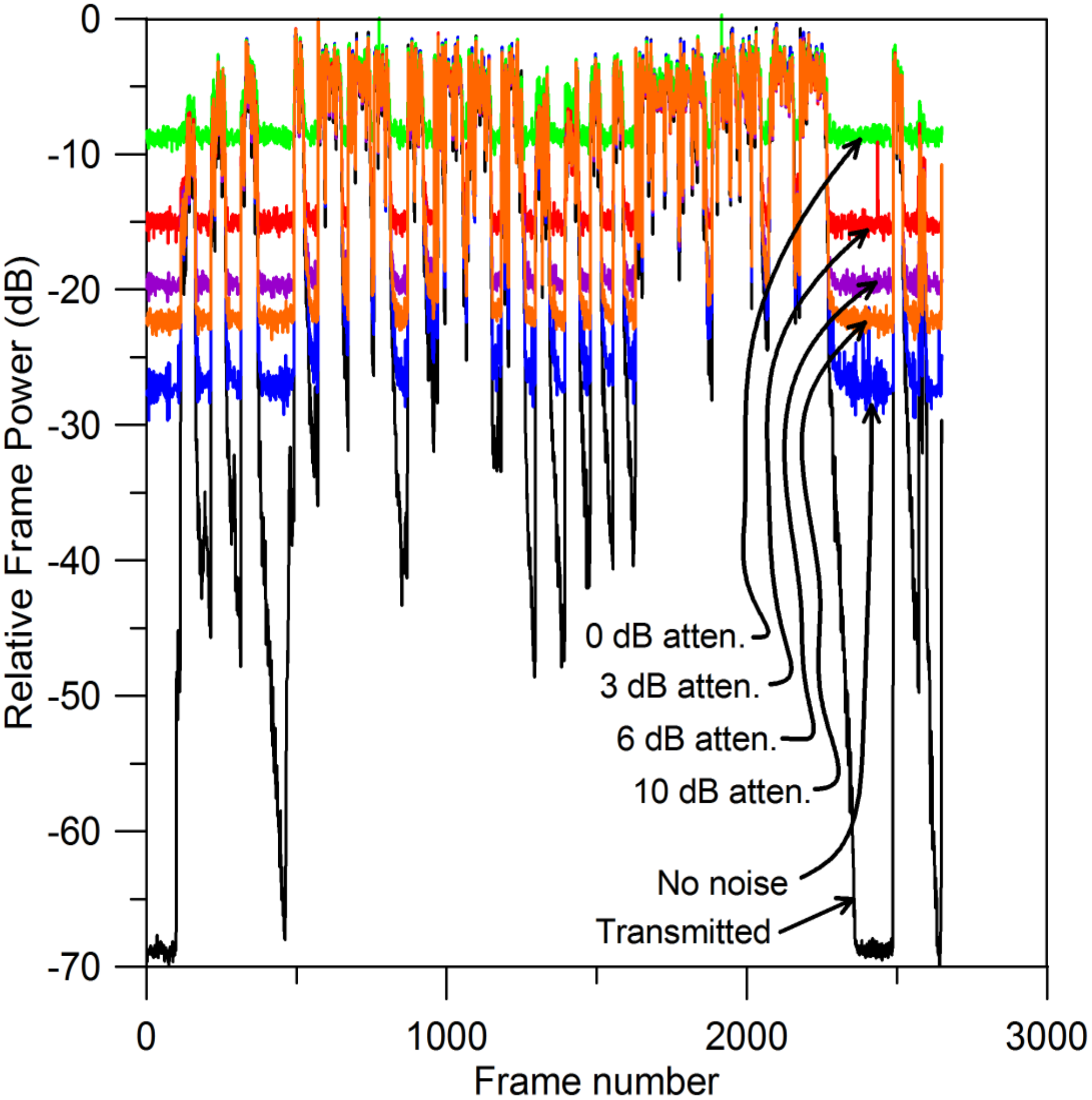}}\hspace{5mm}
\scalebox{.17}{\includegraphics*{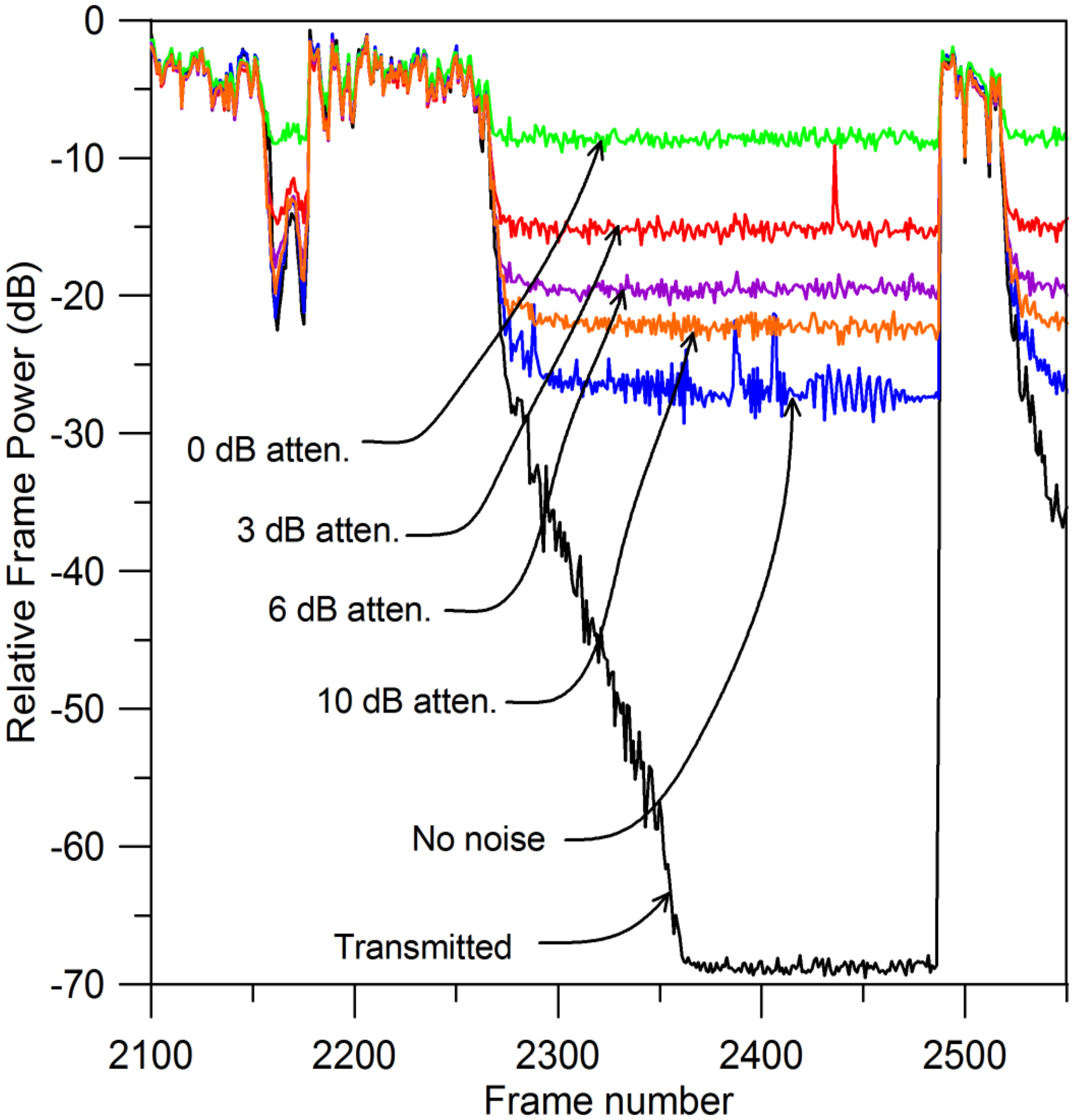}}\\
\vspace*{-2mm}
{\hspace{-1mm}\tiny{(a) \hspace{30mm} (b)}}\\
\vspace*{2mm}
\scalebox{.17}{\includegraphics*{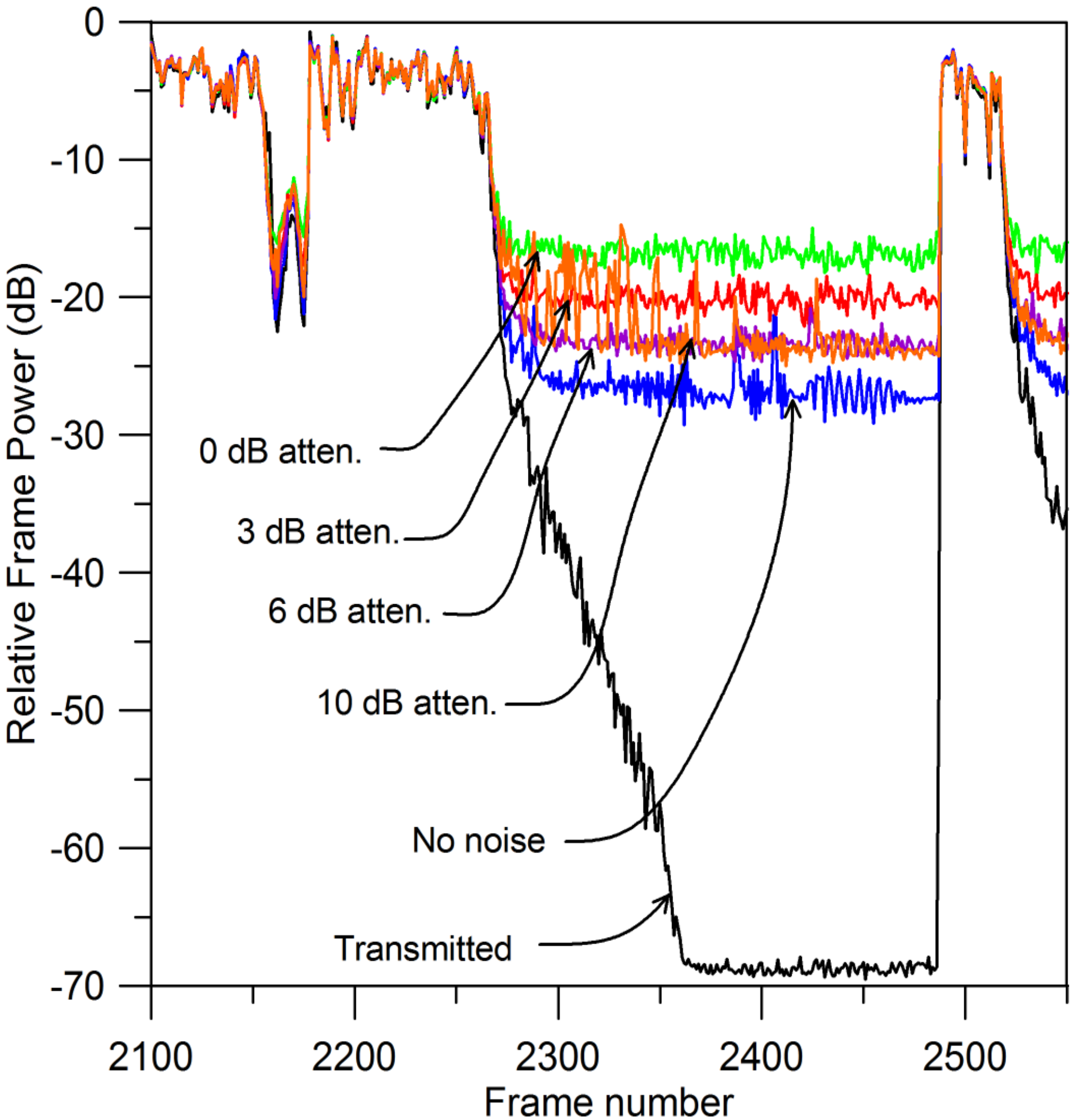}}\hspace{5mm}
\scalebox{.17}{\includegraphics*{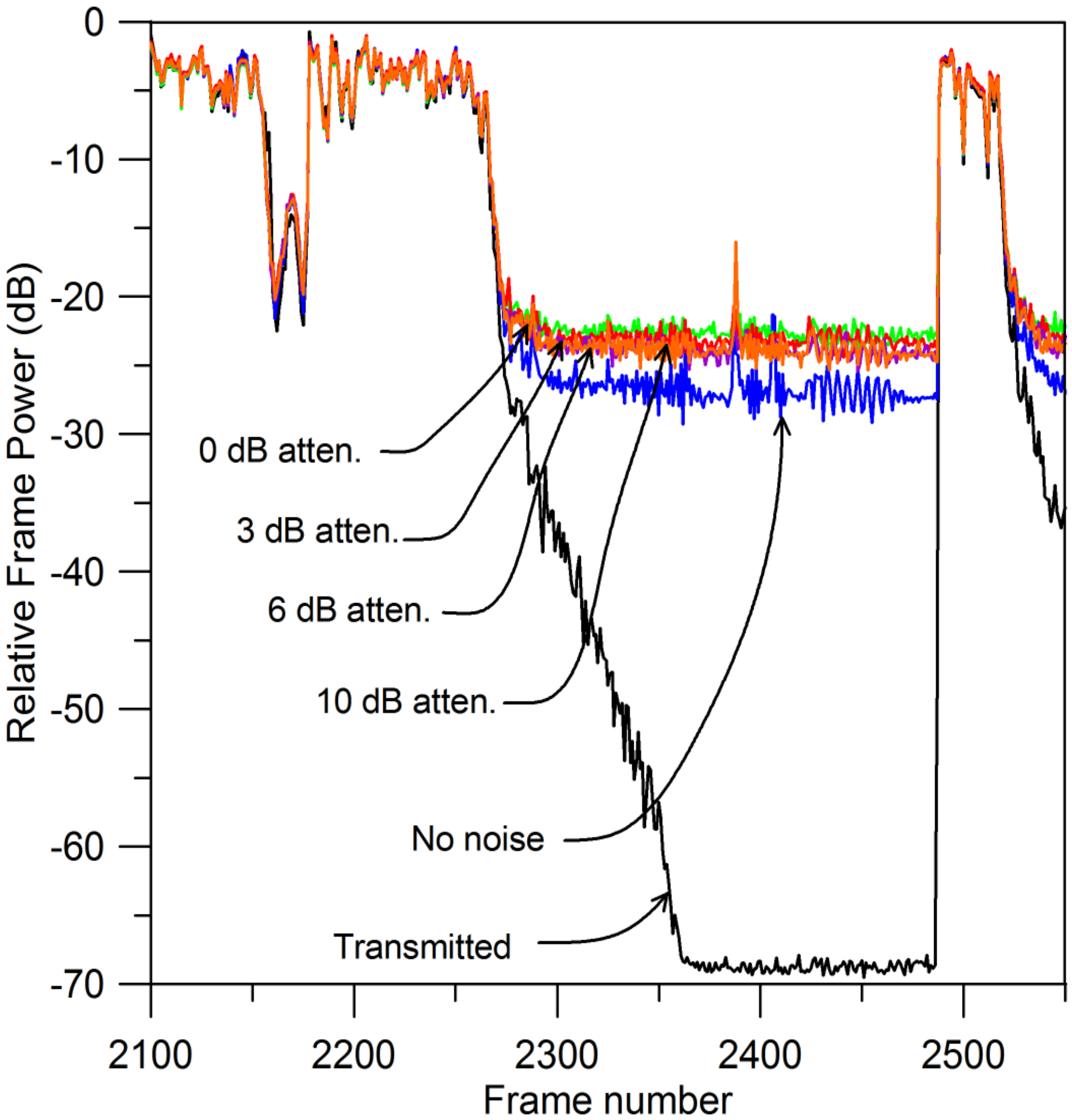}}\\
\vspace*{-2mm}
{\hspace{-1mm}\tiny{(c) \hspace{30mm} (d)}}\\
\caption{Frame power histories: (a) Filter 1 full record; partial records for (b) Filter~1, (c) Filter~2, and (d) Filter~3.}
\label{profile}
\end{figure}

These frame power histories are shown Fig.~\ref{profile}. Fig.~\ref{profile}(a) shows the entire history (approximately 77 seconds) for Filter~1. The effect of the noise is easily seen by viewing a portion of the history as in Figs.~\ref{profile}(b)-\ref{profile}(d).
Each of these shows approximately 450 frames (10 seconds) for one of the three filters.
These figures show how a gap in the vocal signal provides an opportunity to measure noise levels.
Thanks to normalization, the histories are very similar when the vocal signal is present (e.g. frames 2200 to 2250).
But when the vocal signal is absent (e.g. frames 2400 to 2450), the noise induced into the received audio by the interfering RF signal is clearly evident. The figures show that, as expected, lower levels of interfering RF noise power produce lower levels of noise in the received audio.
They also show that the effect of interfering RF power is strongest in the case of Filter 1 and weakest in the case of Filter 3.
That is, the noise that is blue-shifted from the carrier (Filter~1) has the most effect. This is consistent with the noise experiments in other studies, where it was shown that blue-shifted noise has the strongest effect on E-field strength measurements performed with the EIT/AT approach \cite{r20}.

\begin{figure}
\centering
\scalebox{.2}{\includegraphics*{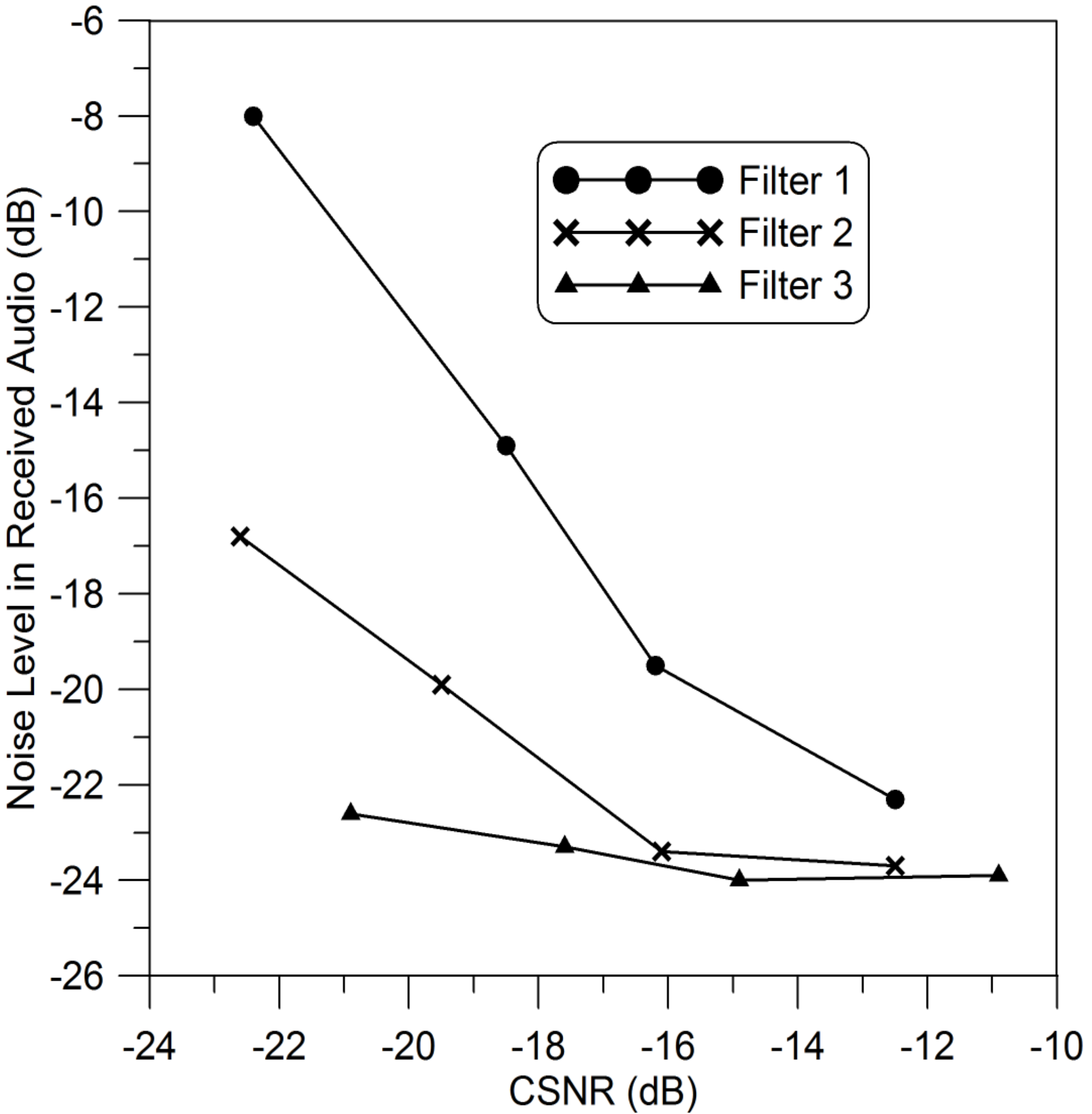}}
\caption{Audio noise levels (relative to the peak signal level) versus CSNR values.}
\label{AudioNoise_vs_CSNR}
\end{figure}

While Fig.~\ref{profile} provides intuitive and accessible demonstrations of these effects, more quantitative results are provided in Table~\ref{t2} and the associated Fig.~\ref{AudioNoise_vs_CSNR}.
To obtain these results we averaged frame powers over 200 frames (approximately 4~seconds) where no vocal signal was present.
These average audio noise frame power values are relative to the peak level of the vocal signal combined with the noise.
Using the fact that the vocal signal power and the noise power are additive, we then adjusted each average audio noise frame power value to report audio noise level in dB relative to the peak signal level.
These audio noise levels, along with the corresponding CSNR values, are given in Table~\ref{t2} and Fig.~\ref{AudioNoise_vs_CSNR}.

When no interfering RF noise is presented at the vapor cell, the audio noise level is 27 dB down.
This noise is audible in a quiet listening environment but would be inaudible in a typical office, retail, or automotive environment.
As more interfering RF noise is introduced, the measured audio noise level increases and the perceived severity of that noise increases accordingly.
For this specific signal and noise combination, noise levels around -$20$ dB may be perceived as and described as ``moderate,'' while levels around -$10$ dB would likely be described as ``severe.''
Fig.~\ref{AudioNoise_vs_CSNR} shows quantitatively how, given a fixed CNSR, noise from filter 1 is much more detrimental than noise from filter 2, which in turn is slightly more detrimental than filter 3.  In addition Fig.~\ref{AudioNoise_vs_CSNR} suggests the presence of an inflection point, perhaps around -15 dB CNSR.
It appears that below this point, changes in CSNR have greater influence in received audio noise level while above this point changes in CSNR have a lesser influence in received audio noise level.

%

\begingroup
\begin{table}
\caption{Calculated CSNR values (ratio of the CW power in the carrier to integrated noise power, both measured at the input to the horn antenna) and noise levels in received audio signal (mean noise level relative to peak signal level.). Note that the labels for the attenuators (i.e., 3~dB, 6~dB, and 10~dB) are only approximate, and we used measured values to calculate CRNR shown in the Table.}
\label{t2}
\begin{center}
\begin{tabular}{|c|c|c|c|}
  \hline

  \hline
  & CSNR (linear / dB) & Audio Noise Level  \\
  \hline
Transmitted signal & --  & $-68.8$~dB \\
Received: No noise & --  & $-27.0$~dB  \\
\hline
\multicolumn{3}{|c|}{Noise conditions}\\
\hline
Filter 1: 10~dB atten. & 0.056~/~$-12.5$~dB   & $-22.3$~dB  \\
Filter 1: 6~dB atten. & 0.024~/~$-16.2$~dB   &$ -19.5$~dB  \\
Filter 1: 3~dB atten. & 0.014~/~$-18.5$~dB   & $-14.9$~dB  \\
Filter 1: 0~dB atten. & 0.0058~/~$-22.4$~dB   & $-8.0$~dB  \\
  \hline
Filter 2: 10~dB atten. & 0.056~/~$-12.5$~dB   & $-23.7$~dB  \\
Filter 2: 6~dB atten. & 0.025~/~$-16.1$~dB   & $-23.4$~dB  \\
Filter 2: 3~dB atten. & 0.011~/~$-19.5$~dB   & $-19.9$~dB  \\
Filter 2: 0~dB atten. & 0.0055~/~$-22.6$~dB   & $-16.8$~dB  \\
  \hline
Filter 3: 10~dB atten. & 0.081~/~$-10.9$~dB   & $-23.9$~dB  \\
Filter 3: 6~dB atten. & 0.032~/~$-14.9$~dB   & $-24.0$~dB  \\
Filter 3: 3~dB atten. & 0.017~/~$-17.6$~dB   & $-23.3$~dB  \\
Filter 3: 0~dB atten. & 0.0081~/~$-20.9$~dB   & $-22.6$~dB  \\
  \hline
  \end{tabular}
\end{center}
\end{table}
\endgroup

We have continued these noise studies by transmitting and receiving a 511-bit pseudo-random bit stream using both AM and FM signals.  These results are presented in a separate publication \cite{biterror}, where we show bit-error-rates for various data rates and for various BLWGN levels.  The results in \cite{biterror} show that while noise is observed in the data, the signal quantity of the received signal is immune to the noise even for low values of CSNR (high noise levels) for the most part. The study in \cite{biterror} also shows that the Rydberg-atom receiver has a bandwidth of about 1~MHz-to-5~MHz (which is independent on the carrier frequency, hundreds of MHz to 1~THz).  This bandwidth limit is due to the time required to populate the atoms to a Rydberg state \cite{biterror}.


\section{Conclusion and Discussion}

Rydberg-atom based receivers/antennas are a new area of research and these types of detection schemes potentially have many advantages over conventional receiving and detection technologies. In this paper we discuss the ability of the atom-based technologies to receive multiple-channels simultaneously. We present one realization of this by  using two different atomic species ($^{85}$Rb and $^{133}$Cs) simultaneously to receive a stereo musical composition using both AM and FM schemes. The output heard from the speakers  had a small amount of noise but was of high-fidelity, in that the musical composition was clearly audible and very understandable. We also investigated the effects of BLWGN on the ability to receive these AM/FM signals.  A general comment about all the results of the atom-based receiver is that the received signals are not noticeably distorted by the various noise levels (i.e., even for CSNR as low as -22~dB). The BLWGN does not cause distortion, but it can cause very audible noise in the received audio signal, depending on the CSNR.  In effect, the atoms act as a filter for the noise.  We demonstrated the long term stability of the approach by streaming an internet radio station and listen to music for several hours at a time as we worked in the laboratory.

This type of atom-based receiver/antennena potentially has several advantages. Most noticeable are; (1) the atoms perform the demodulation and allow for direct read-out of the base-band signal, (2) they allow for multi-band (multi-channel) receiving in one sensor head, (3) one sensor head can operate from hundreds of MHz to 1~THz, and (4) the bandwidth of operation is limited to around 5~MHz, but this bandwidth is constant over the entire frequency range of hundreds of MHz to 1~THz.

While more research is needed to fully understand the {\it pros} and {\it cons} of this approach, the study reported here illustrates the capability of a Rydberg atom-based receiver/antenna system. Furthermore, while it is unclear if the Rydberg-atom approach has advantages over conventional radio technologies and what those advantages might me, the study presented here, and by others, allow us to get a step closer to answering these questions. With all that said, the results in this work do show very interesting applications of ``atomic physics'' applied to a the age old topic of ``radio'' reception: i.e., {\it quantum physics meets radio} (or the {\it atom-radio} as coined in \cite{rc3} and \cite{atomradio}).



\begin{thebibliography}{999}

\bibitem{gal} T.F. Gallagher, {\bf Rydberg Atoms}. Cambridge Univer. Press:Cambridge, 1994.

\bibitem{r2} C.L. Holloway, M.T. Simons, J.A. Gordon, P.F. Wilson, C.M. Cooke, D.A. Anderson, and G. Raithel, {\it IEEE Trans. on Electromagnetic Compat.,} vol. 59, no. 2, 717-728, 2017.
\bibitem{r3} C.L. Holloway, J.A. Gordon, A. Schwarzkopf, D.A. Anderson, S.A. Miller, N. Thaicharoen, and G. Raithel, {\it IEEE Trans. on Antenna and Propag.,} vol. 62, no. 12, 6169-6182, 2014.
\bibitem{r4} J.A. Sedlacek, A. Schwettmann, H. Kubler, R. Low, T. Pfau and J.P. Shaffer, {\it Nature Phys.}, vol. 8, 819, 2012.
\bibitem{r5} C.L. Holloway, J.A. Gordon, A. Schwarzkopf, D.A. Anderson, S.A. Miller, N. Thaicharoen, and G. Raithel, {\it Applied Phys. Lett.,} vol. 104, 244102-1-5, 2014.
\bibitem{r6} J.A. Sedlacek, A. Schwettmann, H. Kubler, and J.P. Shaffer, {\it Phys. Rev. Lett.,} vol. 111, 063001, 2013.
\bibitem{r7} J. A. Gordon, C. L. Holloway, A. Schwarzkop, D. A. Anderson, S. Miller, N. Thaicharoen, G. Raithel, {\it Applied Physics Letters,} vol. 105, 024104, 2014.
\bibitem{r8} H. Fan, S. Kumar, J. Sedlacek, H. Kubler, S. Karimkashi and J.P Shaffer, {\it J. Phys. B: At. Mol. Opt. Phys.,} vol. 48, 202001, 2015.
\bibitem{r9} M. Tanasittikosol, J.D. Pritchard, D. Maxwell, A. Gauguet, K.J. Weatherill, R.M. Potvliege and C.S. Adams, {\it J. Phys B,} vol. 44, 184020, 2011.
\bibitem{r10} C.G. Wade, N. Sibalic, N.R. de Melo, J.M. Kondo, C.S. Adams, and K.J. Weatherill, {\it Nature Photonics}, vol. 11, 40-43, 2017.
\bibitem{r11} H. Fan, S. Kumar, J. Sedlacek, H. Kubler, S. Karimkashi and J.P Shaffer, {\it J. Phys. B: At. Mol. Opt. Phys.,} 48, 202001, 2015.
\bibitem{r12} D.A. Anderson, S.A. Miller, G. Raithel, J.A. Gordon, M.L. Butler, and C.L. Holloway, {\it Physical Review Applied,} 5, 034003, 2016.
\bibitem{r13} D.A. Anderson, S.A. Miller, A. Schwarzkopf, C.L. Holloway, J.A. Gordon, N. Thaicharoen, and G. Raithelet, {\it Physical Review A}, vol. 90, 043419, 2014.
\bibitem{r14} A.K. Mohapatra, T.R. Jackson, and C.S. Adams,  {\it Phys. Rev. Lett.,} vol. 98, 113003, 2007.
\bibitem{r15} C.L. Holloway, M.T. Simons, J.A. Gordon, A. Dienstfrey, D.A. Anderson, and G. Raithel, {\it J. of Applied Physics,} vol. 121, 233106-1-9, 2017.
\bibitem{r16} M.T. Simons,  J.A. Gordon,  and C.L. Holloway, {\it Applied Physics Letters}, vol. 108 174101, 2016.
\bibitem{r17} M.T. Simons,  J.A. Gordon,  and C.L. Holloway, {\it J. Appl. Phys.,} vol. 102, 123103, 2016.
\bibitem{r18} C.L. Holloway, M.T. Simons, M.D. Kautz, A.H. Haddab, J.A. Gordon, T.P. Crowley, {\it Applied Phys. Letters,} vol. 113, 094101, 2018.
\bibitem{r19} M.T. Simons, J.A. Gordon, and C.L. Holloway, {\it Applied Optics,} vol. 57, no. 22, pp. 6456-6460, 2018.
\bibitem{r20} M.T. Simons, M.D. Kautz, C.L. Holloway, D. A. Anderson, and G. Raithel, {\it J. of Applied Physics}, 123, 203105, 2018.

\bibitem{dan} D. Stack, B. Rodenburg, S. Pappas, W. Su, M. St. John, P. Kunz, M. Simons, J. Gordon, C.L. Holloway, ``Rydberg Dipole Antennas'', APS DAMOP, June 5-9, Sacramento, CA, May 2017.

\bibitem{rc1}  D.H. Meyer, K.C. Cox, F.K. Fatemi, and P.D. Kunz, ``Digital communication with Rydberg atoms and amplitude-modulated microwave fields'', {\it Appl. Phys. Lett.}, vol. 12, 211108, 2018.

\bibitem{rc2} K.C. Cox, D.H. Meyer, F.K. Fatemi, and P.D. Kunz, ``Quantum-Limited Atomic Receiver in the Electrically Small Regime'',  arXiv:1805.09808v2, June 19, 2018.

 \bibitem{biterror}    M.T. Simons, A.H. Haddab, R. Horansky, and C.L. Holloway, ``Atom-based receiver: A study of bit-error for a pseudo-random bit stream in the presence of noise'', submitted to {\it Appl. Phys Letter.} 2018.

\bibitem{rc3} D.A. Anderson, R.E. Sapiro, and G. Raithel, ``An atomic receiver for AM and FM radio communication'', arXiv:1808.08589v1, Aug. 26, 2018.

\bibitem{rc4} Z. Song, W. Zhang, H. Liu, X. Liu, H. Zou, J. Zhang, and J. Qu, ``The credibility of Rydberg atom based digital communication over a continuously tunable radio-frequency carrier'',  arXiv:1808.10839v2, Sept., 2018.

\bibitem{atomradio} ``Get ready for atom radio'', {\it MIT Technology Review}, June, 2018: https://www.technologyreview.com/s/611977/get-ready-for-atomic-radio/

\bibitem{chu} L.J. Chu, ``Physical limitations of omni-directional antennas'', {\it J. of Applied Physics,} vol. 19, 1163, 1948.

\bibitem{stackrb} D. A. Steck, ``Rubidium 85 D line data'', revision 2.1.6 Sep. 20, 2013 [Online]. Available: http://steck.us/alkalidata

\bibitem{gal1} W. Li, I. Mourachko, M.W. Noel, and T.F. Gallagher, {\it Phys. Rev. A}, {\bf67}, 052502, 2003.

\bibitem{gal3} M. Mack, F. Karlewski, H. Hattermann, S. H\"{o}ckh, F. Jessen, D. Cano, and J. Fort\'{a}gh,  {\it Phys. Rev. A}, vol. 83, 052515, 2011.

\bibitem{qdcs1} P. Goy, J.M. Raimond, G. Vitrant, and S. Haroche, {\it Physical Review A}, {\bf 26}, 5, Nov. 1982.



\end{thebibliography}
\end{document}